\documentclass[onecolumn,preprintnumbers,amsmath,amssymb,12pt,nofootinbib]{revtex4}

\usepackage{graphicx}
\usepackage{dcolumn}
\usepackage{bm}
\usepackage{dsfont}
\usepackage{fixmath}
\usepackage{amssymb,amsfonts,amsmath}
\usepackage{graphicx}

\newcommand{\dd}{\mathrm{d}}
\newcommand{\ee}{\mathrm{e}}
\newcommand{\ii}{\mathrm{i}}
\newcommand{\R}{\mathds R}
\newcommand{\C}{\mathds C}
\newcommand{\eps}{\varepsilon}

\newcommand{\diff}[2]{\frac{\partial#1}{\partial#2}}
\newcommand{\rW}{\varrho}
\newcommand{\zW}{\zeta}

\newcommand{\Ap}{\mathcal A^+}
\newcommand{\Apm}{\mathcal A^\pm}
\newcommand{\An}{\mathcal A^0}
\newcommand{\Am}{\mathcal A^-}
\newcommand{\Ha}{\mathcal{H}^{(1)}}
\newcommand{\Hb}{\mathcal{H}^{(2)}}
\newcommand{\La}{{\mathbold L}^{(1)}}
\newcommand{\Lb}{{\mathbold L}^{(2)}}
\newcommand{\Lp}{{\mathbold L}^{+}}
\newcommand{\Lm}{{\mathbold L}^{-}}
\newcommand{\Ln}{{\mathbold L}^{0}}
\newcommand{\LI}{{\mathbold L}^{I}}
\newcommand{\Eins}{{\mathbf 1}}
\newcommand{\Fa}{{\mathbold F}_1}
\newcommand{\Fb}{{\mathbold F}_2}
\newcommand{\Fc}{{\mathbold F}_3}
\newcommand{\Fd}{{\mathbold F}_4}
\newcommand{\Rp}{{\mathbold R}^{+}}

\newcommand{\rp}{{\mathbold r}^{+}}

\newcommand{\bPhi}{{\mathbold\Phi}}
\newcommand{\bP}{{\mathbold P}}
\newcommand{\bQ}{{\mathbold Q}}
\newcommand{\TOm}{{\mathbold T}_\Omega}
\normalsize

\begin{document}

\title{\Large Stationary two-black-hole configurations:\\ A non-existence proof}

\author{\bf Gernot Neugebauer}
\email{G.Neugebauer@tpi.uni-jena.de}
\affiliation{Theoretisch-Physikalisches Institut,
          Friedrich-Schiller-Universit\"at,
          Max-Wien-Platz 1,
          07743 Jena, Germany}
\author{\bf J\"org Hennig}
\email{jhennig@maths.otago.ac.nz}
\affiliation{Department of Mathematics and Statistics,
           University of Otago,
           P.O. Box 56, Dunedin 9054, New Zealand}


\begin{abstract}
 Based on the solution of a boundary problem for disconnected (Killing) horizons and the resulting violation of characteristic black hole properties, we present a non-existence proof for equilibrium configurations consisting of two aligned rotating black holes. Our discussion is principally aimed at developing the ideas of the proof and summarizing the results of two preceding papers (Neugebauer and Hennig, 2009 \cite{Neugebauer2009}, Hennig and Neugebauer, 2011 \cite{Hennig2011}). From a mathematical point of view, this paper is a further example (Meinel et~al., 2008 \cite{Meinel2008}) for the application of the inverse (``scattering'') method to a non-linear elliptic differential equation.
\end{abstract}


\maketitle
\section{Introduction \label{sec:Intro}}
The investigation of gravitational interactions in static two-body systems dates back to the early days of General Relativity and was initiated by Hermann Weyl and Rudolf Bach. In a joint paper \cite{Bach} they discussed, as a characteristic example, an axisymmetric configuration consisting of two ``sphere-like'' bodies at rest. Bach, who constructed a corresponding solution for the vacuum region outside the bodies by superposition of two exterior Schwarzschild solutions, noted that this solution becomes singular on the portion of the symmetry axis between the two bodies as expected. In a supplement to Bach's contribution, Weyl focused on the interpretation of this type of singularity and used stress components of the energy-momentum tensor to define a \emph{non-gravitational} repulsion between the bodies which compensates the gravitational attraction. Weyl's result is based on some artificial assumptions but implies an interesting question: Are there repulsive effects of \emph{gravitational} origin which could counterbalance the omnipresent mass attraction?

Newtonian approximations tell us that the interaction of the angular momenta of rotating bodies (``spin-spin interaction'') could indeed generate repulsive effects. This is a good  motivation to study, in a rigorous way, the equilibrium between two (aligned) rotating black holes with parallel (or anti-parallel) spins as a characteristic example for a stationary two-body problem. In preceding papers \cite{Neugebauer2009,Hennig2011}, which involved degenerate (``extreme'') black holes we came to a negative conclusion. This paper is meant to summarize the steps of this non-existence proof, to point out the main points of the matter and to refer non-specialists to papers dealing with the technical details. It should be noted that a non-existence proof for special symmetric equilibrium configurations by Beig et al. \cite{Beig1,Beig2} is essentially based on symmetry arguments and does not apply to black holes with different horizon areas and angular momenta.

Another aspect with some relevance for the two-black-hole configurations in question is the interpretation of the so-called double-Kerr-NUT solution \cite{Neugebauer1980a,Kramer1980}, a seven parameter solution constructed by a two-fold B\"acklund transformation of Minkowski space. Since a single B\"acklund transformation generates the Kerr-NUT solution that contains, by a special choice of its three parameters, the stationary black hole solution (Kerr solution) and since B\"acklund transformations act as a non-linear superposition principle, the double-Kerr-NUT solution was considered to be a good candidate for the solution of the two-horizon problem in question and extensively discussed in the literature \cite{Dietz,Hoenselaers1983,Hoenselaers1984,Kihara1982,Kramer1980,Kramer1986,Krenzer,Manko2000,Manko2001,Tomimatsu,Yamazaki}. However, there was no argument that this particular solution is the \emph{only} candidate. Therefore, defects of this special solution would not \emph{a priori} imply a general non-existence proof for our stationary two-black-hole problem. 

In papers \cite{Neugebauer2000,Neugebauer2003,Neugebauer2009,Hennig2011} we could remove this objection and show that the discussion of a boundary value problem for the Ernst equation which represents a part of the vacuum Einstein equations, necessarily leads to a subclass of the double-Kerr-NUT solution. This result is in line with a theorem of Varzugin \cite{Varzugin1997,Varzugin1998} which says that the $2N$-soliton solution by Belinski and Zakharov contains all possible solutions (if any exist) corresponding to an equilibrium configuration of black holes.

\begin{figure}\centering
 \includegraphics[scale=1]{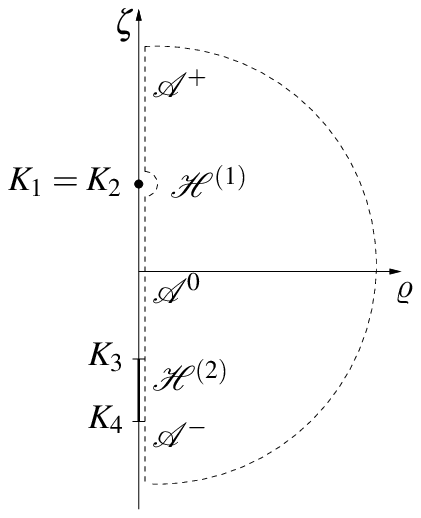}
 \caption{\label{Fig1} Illustration of a two-black-hole configuration with one degenerate (point-like) horizon $\Ha$ ($\rW=0, \zW=K_1$) and one sub-extremal horizon $\Hb$ ($\rW=0, K_3\ge\zW\ge K_4$) in Weyl-Lewis-Papapetrou coordinates. $\Ap$, $\An$ and $\Am$ denote the three parts of the axis of symmetry.}
\end{figure} 

The subclass is characterized by a set of restrictions for the parameters of the general double-Kerr-NUT solution. These restrictions ensure the ``correct'' behavior of the double-Kerr-NUT solution along the axis of symmetry and the horizons. For solving the restrictions we could go back to the already mentioned discussions of the equilibrium conditions for the double-Kerr-NUT solution. After a too restrictive ansatz in \cite{Kramer1980}, Tomimatsu and Kihara \cite{Tomimatsu,Kihara1982} derived and discussed a complete set of equilibrium conditions on the axis of symmetry. Reformulations and numerical studies by Hoenselaers \cite{Hoenselaers1984} made plausible that the double-Kerr-NUT solution cannot describe a two-black-hole equilibrium if the individual Komar masses of its two sources (``horizons'') are assumed to be positive. Manko et al. \cite{Manko2000} and finally Manko and Ruiz \cite{Manko2001} were able to prove the conjecture. The critical point of this proof in view of a non-existence theorem is, however, the presumed positiveness of the two Komar masses. To the best of our knowledge there is no argument in favor of this assumption (on the contrary, Ansorg and Petroff \cite{Ansorg2006} gave convincing counterexamples). Instead, we replace the Komar mass inequality $M_i>0$  for each black hole  ($i=1,2$) by inequalities connecting angular momenta $J_i$ and surface areas $A_i$ ($8\pi|J_i|<A_i,\ i=1,2$) \cite{Hennig2008a}. These relations are based on the causal structure of trapped surfaces in the interior vicinity of any event horizon \cite{Booth}. In the case of two \emph{non-degenerate} black holes it turns out that one of the two inequalities is always violated, i.e. one of the sources cannot be a black hole.

This type of non-existence proof avoids more laborious investigations of the domain off the axis of symmetry. In some degenerate cases, see Fig.~\ref{Fig1}, we need additional eliminating criteria such as the positiveness of the total (ADM) mass or the absence of singular rings for proving the non-existence. We summarize the results of the discussions of all subcases in Sec.~\ref{insi}.

\section{A boundary problem for disconnected horizons}

The exterior vacuum gravitational field of axially symmetric and stationary gravitational sources can be described in cylindrical Weyl-Lewis-Papapetrou coordinates $(\rW, \zW, \varphi, t)$\footnote{
In the following, we also use the complex coordinates $z=\rW+\ii\zW$ and $\bar z = \rW-\ii\zW$. $t$ is the time coordinate.}, in which the line element takes the form
\begin{equation}\label{LE}
 \dd s^2 = \ee^{-2U}\big[\ee^{2k}(\dd\rW^2+\dd\zW^2)
           +\rW^2\dd\varphi^2\big]
           -\ee^{2U}(\dd t+a\,\dd\varphi)^2,
\end{equation}
where the ``Newtonian'' gravitational potential $U$, the gravitomagnetic potential $a$ and the ``superpotential'' $k$ are functions of $\rW$ and $\zW$ alone. At large distances $r=|\sqrt{\rW^2+\zW^2}|\to\infty$ from isolated sources located around the origin of the coordinate system, $r=0$, the spacetime has to be Minkowskian,
\begin{equation}\label{asflat}
 r\to\infty:\quad \dd s^2=\dd\rW^2+\dd\zW^2+\rW^2\dd\varphi^2-\dd t^2.
\end{equation}
According to the objective of this paper (see Fig.~\ref{Fig1}) we will exclusively discuss gravitational fields \eqref{LE} under condition \eqref{asflat}. 

Metric \eqref{LE} admits an Abelian group of motions $G_2$ with the generators (Killing vectors)
\begin{equation}
\begin{aligned}
 &\xi^i=\delta^i_t       && \quad\textrm{(stationarity)}\\
 &\eta^i=\delta^i_\varphi&& \quad\textrm{(axisymmetry)}
\end{aligned}
\end{equation}
where the Kronecker symbols $\delta^i_t$, $\delta^i_\varphi$ indicate that $\xi^i$ has only a $t$-component whereas $\eta^i$ points in the azimuthal ($\varphi$) direction. $\eta^i$ has closed compact trajectories about the axis of symmetry and is therefore space-like off the axis (and the horizons). $\xi^i$ is time-like sufficiently far from the black holes but can become space-like inside ergoregions.  Obviously,
\begin{equation}\label{coordfree}
 \ee^{2U}=-\xi^i\xi_i,\quad a=-\ee^{-2U}\eta^i\xi_i
\end{equation}
is a coordinate-free representation of the gravitational potentials $U$ and $a$.

According to Carter's theorems \cite{Carter} we can assume that the event horizons of the two black holes under discussion are Killing horizons. Here a Killing horizon can be defined by a linear combination $\xi'$ of the Killing vectors $\xi$ and $\eta$,
\begin{equation}\label{KV}
 \xi'=\xi+\Omega\eta
\end{equation}
with the norm
\begin{equation}\label{Vdef}
 \ee^{2V}=-(\xi',\xi')=\ee^{2U}\left[(1+\Omega a)^2-\rW^2\Omega^2\ee^{-4U}\right]
\end{equation}
where $\Omega$ is the constant angular velocity of the horizon. A connected component of the set of points with $\ee^{2V}=0$, which is a null hypersurface, $(\dd\ee^{2V},\dd\ee^{2V})=0$, is called a Killing horizon $\mathcal H(\xi')$,
\begin{equation}\label{KH}
  \mathcal H(\xi'):\quad \ee^{2V}=-(\xi',\xi')=0,\quad (\dd\ee^{2V},\dd\ee^{2V})=0.
 \end{equation}
 
 Since the Lie derivative $\mathcal L_{\xi'}$ of $\ee^{2V}$ vanishes, we have $(\xi',\dd\ee^{2V})=0$. Being null vectors on $\mathcal H(\xi')$, $\xi'$ and $\dd\ee^{2V}$ are proportional to each other,
\begin{equation}\label{kappa1}
 \mathcal H(\xi'):\quad \dd\ee^{2V}=-2\kappa \xi'.
\end{equation}
Using the field equations one can show that the \emph{surface gravity} $\kappa$ is a constant on $\mathcal H(\xi')$.

In the $\rW$-$\zW$ plane ($t=\textrm{constant}$, $\varphi=\textrm{constant}$) of the Weyl-Lewis-Papapetrou coordinate system \eqref{LE} horizons are located on the $\zW$-axis ($\rW=0$) and cover a finite portion of the axis ($\Hb$ in Fig.~\ref{Fig1}) or shrink to a single point ($\Ha$ in Fig.~\ref{Fig1})\cite{Carter}. It turns out that extended horizons (``sub-extremal horizons'') and point-like horizons (``degenerate horizons'') require different considerations. Note that a Killing horizon is always a two-surface in the time slice $t=\textrm{constant}$. The degeneracy to a line or a point is a peculiarity of the special coordinate system.

The dashed line in Fig.~\ref{Fig1} sketches the boundaries of the vacuum region: $\Ap$, $\An$, $\Am$ are the regular parts of the $\zW$-axis (axis of symmetry), $\Ha$ and $\Hb$ denote the two Killing horizons (Fig.~\ref{Fig1} shows a point-like and an extended horizon), and $\mathcal C$ stands for spatial infinity.

The gravitational fields $a$, $k$, $U$ have to satisfy the following boundary conditions
\begin{eqnarray}
 \label{B1}
 \mathcal A^\pm,\An: && \quad a=0,\quad k=0,\\
 \label{B2}
 \mathcal H^{(i)}:       && \quad 1+\Omega_i a=0,\quad i=1,2,\\
 \label{B3}
 \mathcal C:         && \quad U\to 0,\quad a\to 0,\quad k\to 0,         
\end{eqnarray}
where $\Omega_1$ and $\Omega_2$ are the angular velocities of the two
horizons. Equations \eqref{B1} characterize the axis of symmetry (rotation axis). The first relation originates from the second equation in \eqref{coordfree}, since the compact trajectories of $\eta$ with the standard periodicity $2\pi$ become infinitesimal circles with the consequence $\eta\to 0$. The second relation is a necessary condition for elementary flatness (Lorentzian geometry in the vicinity of the rotation axis). Equation \eqref{B2} is a reformulation of Eqs.~\eqref{KH} ($\ee^{2V}=0$) and \eqref{Vdef} since the horizons are located on the $\zW$-axis ($\rW=0$); see Fig.~\ref{Fig1}. Finally, Eq.~\eqref{B3} ensures the asymptotic flatness of the metric \eqref{LE}; see \eqref{asflat}.

In our discussion we will essentially use the Ernst formulation of the field equations \cite{Ernst}. For this purpose, we introduce the complex \emph{Ernst potential}
\begin{equation}
 f=\ee^{2U}+\ii b,
\end{equation}
where the (real) \emph{twist potential} $b$ is defined by
\begin{equation}\label{a}
 a_{,\rW} = \rW\,\ee^{-4U} b_{,\zW},\qquad
 a_{,\zW} = -\rW\,\ee^{-4U} b_{,\rW}.
\end{equation}
In this formulation, a part of the Einstein vacuum equations is equivalent to the
complex \emph{Ernst equation}
\begin{equation}\label{Ernst}
 (\Re f)\Big(f_{,\rW\rW}+f_{,\zW\zW} +\frac{1}{\rW}f_{,\rW}\Big)
 = f_{,\rW}^2 + f_{,\zW}^2.
\end{equation}
Obviously, the imaginary part of the Ernst equation is nothing but the integrability condition $a_{,\rW\zW}=a_{,\zW\rW}$. On the other hand, the condition $b_{,\rW\zW}=b_{,\zW\rW}$ leads back to the field equation for $a$ as an element of the original Einstein equations. 

The metric potential $k$ can be calculated from $f$ via a line integral,
\begin{equation}
 \label{k1}
 k_{,\rW} = \rW\Big[U_{,\rW}^2-U_{,\zW}^2+\frac{1}{4}\ee^{-4U}
         (b_{,\rW}^2-b_{,\zW}^2)\Big],\quad
 k_{,\zW} = 2\rW\Big[U_{,\rW}U_{,\zW}+\frac{1}{4}\ee^{-4U}
         b_{,\rW}b_{,\zW}\Big].
\end{equation}
The result of the integration does not depend on the path of integration, since \eqref{Ernst} implies $(k_{,\rW})_{,\zW}=(k_{,\zW})_{,\rW}$.

Equations \eqref{Ernst} and \eqref{k1} are completely equivalent to the Einstein vacuum equations. Thus one can first integrate the Ernst equation to obtain $\ee^{2U}$ and $b$ or, alternatively, $\ee^{2U}$ and $a$ (see \eqref{a}), and determine the remaining metric potential $k$ (see \eqref{LE}) by line integration \emph{afterward}. Taking advantage of this circumstance we will first analyze the boundary problem
\begin{eqnarray}
 \label{B1x}
 \mathcal A^\pm,\An: && \quad a=0,\\
 \label{B2x}
 \mathcal H^{(i)}:       && \quad 1+\Omega_i a=0,\quad i=1,2,\\
 \label{B3x}
 \mathcal C:         && \quad U\to 0,\quad a\to 0,         
\end{eqnarray}
for the Ernst equation \eqref{Ernst}. Note that the connection between $b=\Im f$ and $a$ is non-local; see \eqref{a}.

\section{The inverse method}

The inverse (scattering) method, first applied for solving initial (value) problems of special classes of \emph{non}-linear partial differential equations in many areas of physics (such as Korteweg-de Vries equation in hydrodynamics, non-linear Schr\"odinger equation in non-linear optics etc.), is based on the existence of a linear problem (LP), whose integrability condition is equivalent to the non-linear differential equation. Surprisingly, the Ernst equation has an LP. This fact is the background for the rich gain of exact solutions with interesting mathematical properties; see \cite{BelZak,Kramer,Meinel2008,Neugebauer1980a,Neugebauer2003,Kramer1980} and the references therein. However, we cannot expect \emph{a priori} one of these solutions to solve our physical question, and are therefore referred to methods applicable to boundary (value) problems. In a sense, we can try to borrow ideas from the above mentioned analysis of initial (value) problems.

 We use the LP
 \cite{Neugebauer1979,Neugebauer1980b}
 \begin{equation}\label{LP}
 \begin{aligned}
  \bPhi_{,z} & = \left[\left(\begin{array}{cc}
                   N & 0\\
                   0 & M\end{array}\right)
                   +\lambda\left(\begin{array}{cc}
                   0 & N\\
                   M & 0\end{array}\right)\right]\bPhi,\\
  \bPhi_{,\bar z} & = \left[\left(\begin{array}{cc}
                   \bar M & 0\\
                   0 & \bar N\end{array}\right)
                   +\frac{1}{\lambda}\left(\begin{array}{cc}
                   0 & \bar M\\
                   \bar N & 0\end{array}\right)\right]\bPhi,
 \end{aligned}                 
 \end{equation}
where the \emph{pseudopotential}
$\bPhi(z,\bar z,\lambda)$ is a $2\times2$ matrix depending on the
spectral parameter
\begin{equation}\label{lambda}
 \lambda=\sqrt{\frac{K-\ii\bar z}{K+\ii z}},\quad K\in\C,
\end{equation}
as well as on the complex coordinates
\begin{equation}
 z=\rW+\ii\zW,\quad \bar z=\rW-\ii\zW,
\end{equation}
whereas $M$, $N$ and the complex conjugate quantities $\bar M$, $\bar N$ are functions of $z$, $\bar z$ (or $\rW$, $\zW$) alone and do not depend on the constant parameter $K$. Since the integrability conditions $\bPhi_{,z\bar z}=\bPhi_{,\bar z z}$ must hold identical in $K$ (or $\lambda$) they yield the first order equations
\begin{equation}\label{AB}
 M_{,\bar z}=M(\bar N-\bar M)-\frac{1}{4\rW}(M+\bar N),\quad
 N_{,\bar z}=N(\bar M-\bar N)-\frac{1}{4\rW}(N+\bar M)
\end{equation}
with the ``first integrals''
\begin{equation}\label{firstin}
 M=\frac{f_{,z}}{f+\bar f},\qquad
 N=\frac{\bar f_{,z}}{f+\bar f},
\end{equation}
where $f$ is any complex function of $z$, $\bar z$. Resubstituting $M$ and $N$ in Eqs. \eqref{AB} one obtains the Ernst equation \eqref{Ernst} for $f(z,\bar z)$. Thus the Ernst equation is the integrability condition of the LP \eqref{LP}. Vice versa, the matrix $\bPhi$ calculated from $M$, $N$ does not depend on the path of integration if $f$ is a solution of the Ernst equation.

Without loss of generality the matrix $\bPhi$ may be assumed to have the form
\begin{equation}\label{norm}
 \bPhi=\left(\begin{array}{cc}
              \psi(\rW,\zW,\lambda) & \psi(\rW,\zW,-\lambda)\\
              \chi(\rW,\zW,\lambda) & -\chi(\rW,\zW,-\lambda)
             \end{array}\right).
\end{equation}
Note that both columns are independent solutions of the LP. The particular form of \eqref{norm} is equivalent to
\begin{equation}\label{uplow}
 \bPhi(-\lambda)=\left(\begin{array}{cc} 1 & 0\\ 0 & -1\end{array}\right)
 \bPhi(\lambda)\left(\begin{array}{cc} 0 & 1\\ 1 & 0\end{array}\right).
\end{equation}
Furthermore
\begin{equation}\label{psibar}
 \bar \psi\left(\rW,\zW,\frac{1}{\bar\lambda}\right)=\chi(\rW,\zW,\lambda)
\end{equation}
due to the special structure of the coefficient matrices of the LP.

For $K\to\infty$ and $\lambda\to -1$ the functions $\psi$, $\chi$ may be normalized by
\begin{equation}\label{normali}
 \psi(\rW,\zW,-1)=\chi(\rW,\zW,-1)=1.
\end{equation}
For $K\to\infty$ and $\lambda\to 1$ one finds 
\begin{equation}\label{fLP}
 f(\rW,\zW)=\chi(\rW,\zW,1)\qquad
 \left(\bar f(\rW,\zW)=\psi(\rW,\zW,1)\right)
\end{equation}
as a consequence of the LP \eqref{LP}. Thus one obtains the solution $f$ of the Ernst equation from the pseudopotential $\bPhi$ ($\chi=\Phi_{21}$, $\psi=\Phi_{11}$) for a particular choice of the spectral parameter $\lambda$. Similarly, the gravitomagnetic potential $a$ can be rediscovered at $K\to\infty$ and $\lambda=1$ in the first derivatives of the pseudopotential $\bPhi$ \cite{Neugebauer2000},
\begin{equation}\label{aLP}
 a(\rW,\zW)=-\rW\ee^{-2U}\left.\left(\diff{}{\lambda}[\chi(-\lambda)-\psi(-\lambda)]\right)\right|_{\scriptsize\begin{array}{c}\lambda=1\\ K\to\infty\end{array}}-C.
\end{equation}
To prove this one has to make use of the LP \eqref{LP} and Eqs. \eqref{firstin} and \eqref{a}. The arbitrary real constant $C$ may be fixed by setting $a=0$ for $r=\sqrt{\rW^2+\zW^2}\to\infty$. An alternative form of 
\eqref{aLP} is 
\begin{equation}\label{aLPalt}
 a(\rW,\zW)=\ii\ee^{-2U}\left.\left(K^2\diff{}{K}[\chi(-\lambda)-\psi(-\lambda)]\right)\right|_{\scriptsize\begin{array}{c}\lambda=1\\ K\to\infty\end{array}}-C.
\end{equation}
(Note that $\partial/\partial K=\lambda_{,K}\partial/\partial\lambda$ with $\lambda$ from \eqref{lambda}.) The idea of the inverse (scattering) method is to discuss $\bPhi$, for fixed but arbitrary values of $\rW$, $\zW$ ($z$, $\bar z$) as a holomorphic function of $\lambda$. We will show that the boundary values \eqref{B1}-\eqref{B3} together with the LP \eqref{LP} yield the necessary information that enables us to construct $\bPhi(\rW,\zW,\lambda)$. (The Ernst potential and the gravitomagnetic potential can then be determined in a simple way; see \eqref{fLP}, \eqref{aLP}). To realize the program we will integrate the LP along the dashed line in Fig.~\ref{Fig1} starting from and returning to any point $\rW=0$, $\zW\in\Ap$. Obviously, $\lambda$ degenerates at $\rW=0$ to $\lambda=\pm 1$, for which reason we perform the discussion in the two sheets of the Riemann $K$-surface connected with $\lambda$ according to \eqref{lambda}. Note that the mapping of the Riemann surface of $K$ onto the $\lambda$-plane depends on the parameters $\rW$, $\zW$. Thus one has movable branch points $K_\textrm{B}=\ii\bar z$, $\bar K_\textrm{B}=-\ii z$ and the branch cut between them changes with the coordinates.

\section{Solution of the boundary problem}

\subsection{Integration along the boundary\label{inboun}}

Integrating the LP \eqref{LP} along $\Ap$, $\Ha$, $\An$, $\Hb$, $\Am$ and using \eqref{firstin} one finds for the values  of $\bPhi$ on each interval $I$, 
$I=\Ap, \Ha, \An, \Hb, \Am$, the representation
\begin{equation}\label{phiI}
 I:\quad \bPhi=\left(\begin{array}{cc}\bar f^I & 1\\ f^I & -1\end{array}\right)
               \LI,\quad \LI=
               \left(\begin{array}{cc}A^I(K) & B^I(K)\\ C^I(K) & D^I(K)
                     \end{array}\right),
\end{equation}
where $f^I$ is the value of the Ernst potential in the interval $I$ and $A^I(K)$, $B^I(K)$, $C^I(K)$, $D^I(K)$ are integration ``constants'' depending on $K$ alone. On the axis of symmetry ($I=\Apm,\An$) $f$ is a function of $\zW$ alone ($\rW=0$, $\zW\in\Apm,\An$),
\begin{equation}\label{phiIa}
 I=\Apm,\An:\quad f^I=f^I(\zW).
\end{equation}
The same holds on extended horizons which can be characterized by $\rW=0$, too.
Weyl-Lewis-Papapetrou coordinates are, however, inappropriate for integrating along \emph{degenerate} (``point-like'') horizons. As was shown by Meinel, see \cite{Meinel2008}, this defect can be repaired by the introduction of suitable (local) coordinates in the vicinity of such a horizon. To demonstrate the procedure we consider the point-like horizon in Fig.~\ref{Fig1} at $\zW=K_1$ and replace $\rW$ and $\zW$ by polar coordinates $(R,\theta)$, $\rW=R\sin\theta$, $\zW=K_1+R\cos\theta$, in which the horizon is described by $R\to 0$, $\theta\in[0,\pi]$, i.e. by a line in an $R$-$\theta$ diagram. Performing the integration along the degenerate horizon in these coordinates one indeed obtains the structure
\eqref{phiI} with $f^I=f^I(\theta)$. Thus we have
\begin{equation}
 I=\Ha,\Hb:\quad
   f^I=\begin{cases} f^I(\zW)    & \textrm{for extended horizons}\\
                      f^I(\theta) & \textrm{for point-like horizons}
        \end{cases}.
\end{equation}
The definition of our Killing horizons is based on the Killing vectors $\xi'=\xi+\Omega\eta$, see \eqref{KV}, \eqref{Vdef}, \eqref{KH}, where $\Omega=\Omega_1$ for $\Ha$ and $\Omega=\Omega_2$ for $\Hb$. To exploit the characteristic properties of the horizons such as \eqref{KH} we discuss metric potentials $\ee^{2U'}$, $a'$ constructed from $\xi'$, $\eta'=\eta$ according to 
\eqref{coordfree}. Thus we obtain the transformations
\begin{equation}\label{tran1}
 \ee^{2U'}\equiv\ee^{2V} = \ee^{2U}[(1+\Omega a)^2-\Omega^2\rW^2\ee^{-4U}],
\end{equation}
\begin{equation}\label{tran2}
 (1-\Omega a')\ee^{2U'}=(1+\Omega a)\ee^{2U},
\end{equation}
where $\Omega=\Omega_1$ for $\Ha$ and $\Omega=\Omega_2$ for $\Hb$.  It can easily be verified that the line element retains its form \eqref{LE} after the coordinate transformation
\begin{equation}\label{transf}
 \rW'=\rW,\quad \zW'=\zW,\quad\varphi'=\varphi-\Omega t,\quad t'=t,
\end{equation}
where $\Omega=\Omega_1,\Omega_2$, if one simultaneously replaces $\ee^{2U}$ and $a$ by $\ee^{2U'}$ and $a'$. Because of \eqref{transf} we may call the primed quantities $\ee^{2U'}$, $a'$, etc. ``corotating potentials''. Applying \eqref{a} to the primed potentials $a'$, $b'$ and using \eqref{tran1}, \eqref{tran2}, \eqref{a} one obtains the corotating Ernst potential $f'$ from $f$ and finally the corotating quantities $M'$, $N'$ via \eqref{firstin}. This procedure ensures from the outset that $f'$ satisfies the Ernst equation and guarantees the existence of an LP \eqref{LP} in the corotating system. The $\bPhi$-matrices of the two systems of reference are connected by the relation
\begin{equation}\label{phitran}
 \bPhi'=\TOm\bPhi,
\end{equation}
where
\begin{equation}\label{Ttran}
 \TOm=\left(\begin{array}{cc}
             1+\Omega a-\Omega\rW\ee^{-2U} & 0\\
             0 & 1+\Omega a+\Omega\rW\ee^{-2U}
            \end{array}\right)
  +\ii(K+\ii z)\Omega\ee^{-2U}
   \left(\begin{array}{cc} -1 & -\lambda\\ \lambda & 1\end{array}\right)
\end{equation}
with $\Omega=\Omega_1,\Omega_2$. This can be checked up by a straightforward calculation.

The matrix $\bPhi$ as defined in \eqref{norm} can be considered as a unique function of $\lambda$, which is therefore defined on both sheets of the $K$-surface. From this point of view, Eqs. \eqref{phiI} determine $\bPhi$ on one sheet only, say, on the upper sheet with $\lambda=1$.\footnote{$\lambda=1$ is arbitrarily ascribed to the upper sheet.} Its values on the other (lower) sheet with $\lambda=-1$ result from \eqref{uplow}. Consider now $\bPhi$ along $\mathcal C$. It does not depend on $\rW$, $\zW$, i.e., $\bPhi[\mathcal C]=\bPhi(K)$, since $M$ and $N$ vanish (see \eqref{firstin}, \eqref{B3}). Along $\mathcal C$, the spectral parameter $\lambda$ \eqref{lambda} changes from $\lambda=1$ at $\rW=0$, $\zW\to\infty$ to $\lambda=-1$ at $\rW=0$, $\zW\to-\infty$, i.e., starting in the upper sheet at the points of intersection $\Ap/\mathcal C$, one arrives in the lower sheet at $\Am/\mathcal C$. (Using polar coordinates $\zW=r\cos\theta$, $\rW=r\sin\theta$, $\theta\in[0,\pi]$, one has $\lambda=\sqrt{(K-r\ee^{\ii\theta})/(K-r\ee^{-\ii\theta})}\to\lambda=\ee^{\ii\theta}$ as $r\to\infty$ and therefore $\lambda=1$ for $\theta=0$ and $\lambda=-1$ for $\theta=\pi$.) Hence, to return to the upper sheet ($\lambda=1$), one has to apply \eqref{uplow} to $\bPhi[\mathcal C]=\bPhi[\Ap/\mathcal C]=\lim_{\zW\to\infty}\bPhi[\Ap]$, where we have used that $\bPhi$ has to be continuous at the point of intersection $\Ap/\mathcal C$. Thus we obtain
\begin{equation}
 \bPhi[\mathcal C/\Am]=\lim\limits_{\zW\to\infty}\bPhi[\Am]
 =\left(\begin{array}{cc} 1 & 0\\ 0 & -1\end{array}\right)
  \left\{\lim\limits_{\zW\to\infty}\bPhi[\Ap]\right\}
  \left(\begin{array}{cc} 0 & 1\\ 1 & 0\end{array}\right),
\end{equation}
and together with \eqref{phiI}
\begin{equation}
 \left(\begin{array}{cc}\label{pmtrans}
        A^-(K) & B^-(K)\\ C^-(K) & D^-(K)\end{array}\right)
         =\left(\begin{array}{cc} 0 & 1\\ 1 & 0\end{array}\right)
  \left(\begin{array}{cc}
    A^+(K) & B^+(K)\\ C^+(K) & D^+(K)
   \end{array}\right)
  \left(\begin{array}{cc} 0 & 1\\ 1 & 0\end{array}\right),
\end{equation}
where $A^-$ means $A^I$ for $I=\Am$, etc.

In a similar way one can use continuity arguments to interlink the $ABCD$-matrices at the other points of intersection. Since the Ernst equation has to hold at all these points, $f$ must be continuous and unique there. Note that the intersection values
\begin{equation}\label{inval}
 f_1=f[\Ap/\Ha],\quad
 f_2=f[\Ha/\An],\quad
 f_3=f[\An/\Hb],\quad
 f_4=f[\Hb,\Am]
\end{equation}
are purely imaginary: As a metric coefficient, $\ee^{2V}=-g'_{tt}$ is continuous at the points of intersection. According to \eqref{tran1} and \eqref{B1} $\ee^{2V}=\ee^{2U}$ on the regular parts of the $\zW$-axis. Since $\ee^{2V}=0$ on the horizons, $\ee^{2U}$ has to  vanish at these points.

Furthermore, the validity of the Ernst equations of the non-rotating and corotating system at the points of intersection implies that $\bPhi$ and $\bPhi'$ must be continuous there as well. By way of example let us consider the point of intersection $\Am/\Hb$ ($\rW=0$, $\zW=K_4$) of Fig.~\ref{Fig1}, i.e., the transition from a regular axis ($\Am$) to an extended horizon ($\Hb$). According to 
\eqref{phiI} and \eqref{inval} one has
\begin{equation}\label{cont1}
 \bPhi[\Am/\Hb]=\left(\begin{array}{cc} -f_4 & 1\\ f_4 & -1\end{array}\right)
 \Lm
 =\left(\begin{array}{cc} -f_4 & 1\\ f_4 & -1\end{array}\right)
 \Lb,
\end{equation}
where $\Lm=\LI$ for $I=\Am$ and $\Lb=\LI$ for $I=\Hb$.
Note that the determinant of the first factor of the matrix products vanishes such that one may cancel one line of the matrix equation. The corresponding equation for $\bPhi'$ at $z=\ii K_4$ follows from Eqs. \eqref{phiI}, \eqref{phitran} and \eqref{Ttran} under conditions \eqref{B1} ($a=0$ on $\Am$) and \eqref{B2} ($1+\Omega_2 a=0$ on $\Hb$)
\begin{equation}\label{cont2}
 \begin{aligned}
 \bPhi[\Am/\Hb]  = \left(\begin{array}{cc} -f_4-2\ii\Omega_2(K-K_4) & 1\\ 
 f_4+2\ii\Omega_2(K-K_4) & -1\end{array}\right)
 \Lm
  = 2\ii\Omega_2(K-K_4)\left(\begin{array}{cc} -1 & 0\\ 1 & 0\end{array}\right) \Lb.
 \end{aligned}
\end{equation}
Again, the matrix equation consists of two identical lines. Combining a line of 
\eqref{cont1} with a line of \eqref{cont2} one obtains
\begin{equation}\label{cont2a}
 \begin{aligned}
 \left(\begin{array}{cc} f_4 & -1\\ f_4+2\ii\Omega_2(K-K_4) & -1\end{array}\right)
 \Lm
  =\left(\begin{array}{cc} f_4 & -1\\ 2\ii\Omega_2(K-K_4) & 0\end{array}\right)
 \Lb.
 \end{aligned}
\end{equation}

Consider now two \emph{extended} horizons: $\Ha: \zW\in[K_1,K_2]$, $\Hb: \zW\in[K_3,K_4]$, $K_1>K_2>K_3>K_4$. Following the idea that led to equation \eqref{cont2a} one can continue the connection of adjoining ${\mathbold L}$-($ABCD$-) matrices. Involving \eqref{pmtrans} and \eqref{cont2a} and defining
\begin{equation}\label{Fi}
  {\mathbold F}_i:=\left(\begin{array}{cc} -f_i & 1\\ -f_i^2 & f_i\end{array}\right),\quad i=1,2,3,4,
\end{equation}
one arrives at the following chain of equations that reflects the integration of the LP along the closed contour $\Ap\Ha\An\Hb\Am\mathcal C\Ap$,
\begin{equation}\label{cont4}
 \begin{aligned}
  &\Lp=\left(\Eins+\frac{\Fa}{2\ii\Omega^{(1)}(K-K_1)}\right)\La,\quad
  \La=\left(\Eins-\frac{\Fb}{2\ii\Omega^{(2)}(K-K_2)}\right)\Ln\\
  &\Ln=\left(\Eins+\frac{\Fc}{2\ii\Omega^{(3)}(K-K_3)}\right)\Lb,\quad
  \Lb=\left(\Eins-\frac{\Fd}{2\ii\Omega^{(4)}(K-K_4)}\right)\Lm,\\
  &\hspace{4cm}\Lm=\left(\begin{array}{cc}0 & 1\\ 1 & 0\end{array}\right)\Lp
      \left(\begin{array}{cc}0 & 1\\ 1 & 0\end{array}\right),
 \end{aligned}
\end{equation}
where $\Omega^{(1)}=\Omega^{(2)}=\Omega_1$, $\Omega^{(3)}=\Omega^{(4)}=\Omega_2$.

\emph{Point-like} horizons can be involved without any difficulty by setting $K_1=K_2$ or/and $K_3=K_4$ in relations \eqref{cont4}. Consider, for example, the point-like horizon in Fig.~\ref{Fig1}. Though the horizon is placed on a point (as a peculiarity of the Weyl-Lewis-Papapetrou coordinate system), the Ernst potential has different values at $\Ap/\Ha$ and $\Ha/\An$: $f_1\neq f_2$; see \eqref{inval}. Setting $\zW=K_1=K_2$ does not require special considerations such  that one can adopt the discussion of the ``extended'' case step by step.

Eliminating $\La$, $\Ln$, $\Lb$, $\Lm$ by means of \eqref{cont4} and defining
\begin{equation}\label{Rpdef}
 \Rp:=\prod\limits_{i=1}^4\left(\Eins-(-1)^i\frac{{\mathbold F}_i}{2\ii\Omega^{(i)}(K-K_i)}\right)\left(\begin{array}{cc}0 & 1\\ 1 & 0\end{array}\right),
\end{equation}
where $\Omega_1=\Omega^{(1)}=\Omega^{(2)}$, $\Omega_2=\Omega^{(3)}=\Omega^{(4)}$,  one arrives at the final result of the integration along the boundaries
\begin{equation}\label{finres}
 \Lp\left(\begin{array}{cc} 0 & 1\\ 1 & 0\end{array}\right)(\Lp)^{-1}=\Rp,
\end{equation}
which specifies the holomorphic structure of the ``integration constants'' $\mathbold L$ as functions of the complex spectral parameter $K$. Since the trace of the left hand side of \eqref{finres} vanishes, $\Rp$ has to be trace free,
\begin{equation}\label{Kcond}
 \mathrm{tr}\,\Rp=0.
\end{equation}
As a condition identical in $K$, equation \eqref{Kcond} yields four constraints among $\Omega_1$, $\Omega_2$; $K_1-K_2$, $K_2-K_3$, $K_3-K_4$; $f_1$, $\dots$, $f_4$. By way of example,
\begin{equation}
 \frac{\Omega_1}{\Omega_2}=\frac{f_1^2-f_2^2}{f_3^2-f_4^2}
\end{equation}
as a consequence of $\lim_{K\to\infty}\mathrm{tr}(\Rp_{,K}K^2)=0$. The asymptotic behavior of $\Lp$ is prescribed by \eqref{norm}, \eqref{normali}, \eqref{fLP} and \eqref{phiI}, where one has to choose $I=\Ap$,
\begin{equation}\label{Linf}
 \Lp =  \left(\begin{array}{cc}
    A^+(K) & B^+(K)\\ C^+(K) & D^+(K)
   \end{array}\right)
    \to \left(\begin{array}{cc} 1 & 0\\ 0 & 1\end{array}\right)\quad\textrm{as}\quad K\to\infty.
\end{equation}
Nevertheless, equation \eqref{finres} is not sufficient for determining $\Lp$ uniquely. To illustrate the degree of freedom, we factorize $\Lp$,
\begin{equation}\label{factor}
 \Lp=\left(\begin{array}{cc}F(K) & 0\\ G(K) & 1\end{array}\right)
  \left(\begin{array}{cc}\alpha(K) & \beta(K)\\ \beta(K) & \alpha(K)\end{array}\right)
\end{equation}
which is always possible for $\det\Lp\neq 0$ and $A^+\neq \pm B^+$; see \eqref{Linf}. Inserting \eqref{factor} into \eqref{finres}, one obtains 
\begin{equation}\label{FG}
 \left(\begin{array}{cc} -G & F\\ \frac{1-G^2}{F} & G\end{array}\right)=\Rp,
\end{equation}
i.e., $F(K)=\bar F(\bar K)$ and $G(K)=-\bar G(\bar K)$ are uniquely determined. Both functions are regular everywhere in the complex $K$-plane with the exception of simple poles (two extended horizons) or/and confluent poles of second order at most (one or two point-like horizons). 

In the next section, we shall determine the axis values of the Ernst potential $f^+$ from $\Rp$ and show that $\alpha$ and $\beta$ do not affect these values, i.e., $\scriptsize\left(\begin{array}{cc}\alpha & \beta\\ \beta & \alpha\end{array}\right)$                                                                                                                                                                                              
 is a gauge matrix.

\subsection{Axis values of the Ernst potential}
The elements of the matrix $\bPhi=(\Phi_{ik})$ \eqref{norm} are unique functions of $\lambda$ in the vacuum region outside the horizons. That implies that its elements $\psi$ and $\chi$ must be unique ($\Phi_{11}=\Phi_{12}$, $\Phi_{21}=-\Phi_{22}$) at the confluent branch points\footnote{$K_\mathrm{B}=\ii\bar z\to K_\mathrm{B}=-\ii z\to\zW$ as $\rW\to0$.} $K_\mathrm{B}=\bar K_\mathrm{B}=\zW$ of the Riemann $K$-surfaces belonging to axis values $\rW=0$, $\zW\in\Apm,\An$, i.e., according to \eqref{phiI}, \eqref{phiIa}
\begin{equation}\label{Achs}
\begin{aligned}
 & I=\Apm,\An;\ K=K_\mathrm{B}=\zW:\quad\\
 & \bPhi=\left(\begin{array}{cc}\psi & \psi\\ \chi & -\chi\end{array}\right)
 =\left(\begin{array}{cc}\bar f^I(\zW) & 1\\f^I(\zW) & -1\end{array}\right)
 \left(\begin{array}{cc}A^I(\zW) & B^I(\zW)\\C^I(\zW) & D^I(\zW)\end{array}\right),\quad
 \bPhi\left(\begin{array}{cc} 0 & 1\\ 1 & 0\end{array}\right)\bPhi^{-1}=
 \left(\begin{array}{cc}1 & 0\\ 0 & -1\end{array}\right).
 \end{aligned}
\end{equation}
Then
\begin{equation}\label{Achs1}
 I=\Apm,\An:\quad f^I(\zW)=\frac{D^I(\zW)+C^I(\zW)}{A^I(\zW)+B^I(\zW)},\quad \bar f^I(\zW)=\frac{D^I(\zW)-C^I(\zW)}{A^I(\zW)-B^I(\zW)},
\end{equation}
i.e. one can express the values of the Ernst potential on the regular portions of the axis by the ``integration constants'' $A$, $B$, $C$, $D$. According to \eqref{factor}, this means for $f^I(\zW)=f^+(\zW)$, $I=\Ap$,
\begin{equation}\label{Achsf}
 f^+(\zW)=\frac{1+G(\zW)}{F(\zW)},\quad
 \bar f^+(\zW)=\frac{1-G(\zW)}{F(\zW)},
\end{equation}
or
\begin{equation}\label{Achsinv}
 F(\zW)=\frac{2}{f^+(\zW)+\bar f^+(\zW)},\quad
 G(\zW)=\frac{f^+(\zW)-\bar f^+(\zW)}{f^+(\zW)+\bar f^+(\zW)},
\end{equation}
i.e., the matrix $\scriptsize\left(\begin{array}{cc}\alpha(\zW) & \beta(\zW)\\ \beta(\zW) & \alpha(\zW)\end{array}\right)$ does not affect the axis values of the Ernst potential on $\Ap$. Because of the successive transformation \eqref{cont4} this holds for $f^-(\zW)$ and $f^0(\zW)$, too. (Note that \eqref{Achs1} and \eqref{pmtrans} imply $f^+(\zW)=1/f^-(\zW)$, $\zW\in\Ap,\Am$, where $f^+$ and $f^-$ are continuations of $f^+(\zW)$, $\zW\in\Ap$, $f^-(\zW)$, $\zW\in\Am$ via $A(K)$, $B(K)$, $C(K)$, $D(K)$.) Thus we obtain the axis values $f^\pm(\zW)$, $f^0(\zW)$ via \eqref{Achsf} and \eqref{FG} from $\Rp(K=\zW)$, see \eqref{Rpdef}, as well-defined functions of $\zW$. The parameters entering these functions are restricted by the constraints $\mathrm{tr}\,\Rp=0$.

Since the matrix $\scriptsize\left(\begin{array}{cc}\alpha(\zW) & \beta(\zW)\\ \beta(\zW) & \alpha(\zW)\end{array}\right)$ does not affect $f^\pm$, $f^0$ we may set $\alpha(\zW)=1$, $\beta(\zW)=0$ with the consequence
\begin{equation}\label{choice}
 \left(\begin{array}{cc}\alpha(K) & \beta(K)\\ \beta(K) & \alpha(K)\end{array}\right)
 =\left(\begin{array}{cc} 1 & 0\\ 0 & 1\end{array}\right).
\end{equation}
(Note that $\alpha(K)$, $\beta(K)$ are analytic continuations of $\alpha(\zW)$, $\beta(\zW)$.) The particular choice \eqref{choice} is closely connected with a gauge transformation of the matrix $\bPhi$: Any transformation
\begin{equation}\label{gtran}
 \bPhi_\mathrm{new}=\bPhi_\mathrm{old}\left(\begin{array}{cc} a(K) & b(K)\\ b(K) & a(K)\end{array}\right)
\end{equation}
with 
\begin{equation}
 a(K)=\overline{a(\bar K)},\quad b(K)=-\overline{b(\bar K)};\quad
 a(K)\to 1,\quad b(K)\to 0\quad\textrm{as}\quad K\to\infty
\end{equation}
leaves the normalizations \eqref{norm}-\eqref{normali} unaffected and can be used to remove the $\alpha$-$\beta$-matrix from \eqref{factor}. Hence \eqref{gtran} is a gauge transformation and one can adjust the gauge so that
\begin{equation}\label{gauge}
 \Lp=\left(\begin{array}{cc} F(K) & 0\\ G(K) & 1\end{array}\right),\quad
 \left(\begin{array}{cc}\alpha(K) & \beta(K)\\ \beta(K) & \alpha(K)\end{array}\right)=\Eins
\end{equation}
in accordance with \eqref{choice}.
By \eqref{phiI} one has
\begin{equation}
 I=\Ap:\quad \Phi_{12}=\psi(-1)=1,\quad \Phi_{22}=-\chi(-1)=1,
\end{equation}
i.e., the gauge \eqref{gauge}  is equivalent to the formulation of special initial conditions for $\psi$ and $\chi$ at some starting point $\rW=0$, $\zW=\zW_0\in\Ap$, $\lambda=-1$ ($K$ in the lower sheet) of the integration along the closed dashed line in Fig.~\ref{Fig1} (which we performed with unspecified integration constants $A^+$, $B^+$, $C^+$, $D^+$).

Let us summarize the results of the integration of the LP along the boundary and the determination of the axis values $f^\pm(\zW)$, $f^0(\zW)$: With the standard gauge \eqref{gauge} and the representation \eqref{phiI} we obtain for $\bPhi$ on $\Ap$
\begin{equation}\label{phimin}
 I=\Ap:\quad \bPhi=\left(\begin{array}{cc}\bar f^+(\zW) & 1\\ f^+(\zW) & -1\end{array}\right)\Lp,\quad
 \Lp=\left(\begin{array}{cc} F(K) & 0\\ G(K) & 1\end{array}\right),
\end{equation}
where $F(K)$, $G(K)$ are elements of the matrix $\Rp$; see \eqref{FG} and \eqref{Rpdef}.

$\Rp$ must be trace free,
\begin{equation}\label{traf}
 \mathrm{tr}\,\Rp=0,
\end{equation}
see \eqref{Kcond}, which affects via $F(K)$ and $B(K)$ the constant parameters entering the Ernst potential $f^+(\zW)$ on the axis $\Ap$
\begin{equation}
 I=\Ap:\quad f^+(\zW)=\frac{1+G(\zW)}{F(\zW)},
\end{equation}
see \eqref{Achsf}. The particular form of the pseudopotential $\bPhi$ on the intervals $\Ha$, $\An$, $\Hb$, $\Am$ and the axis values $f^0(\zW)$, $f^-(\zW)$ result from \eqref{cont4} with the ``starting matrix'' $\Lp$ as chosen in \eqref{gauge} and expressions \eqref{Achs1}.

$f^+(\zW)$ seems to be a quotient of two normalized\footnote{The fourth order coefficient is equal to one.} polynomials of fourth degree; see \eqref{Achsf}, \eqref{FG}, \eqref{Rpdef}. However, the constraints \eqref{Kcond} take care that the numerator as well as the denominator are of second degree. Inserting the first equation in \eqref{Achs} ($I=\Ap$) into the second one and using \eqref{finres} one obtains
\begin{equation}\label{Rpx}
 \Rp(\zW)=
  \left(\begin{array}{cc}\bar f(\zW) & 1\\ f(\zW) & -1\end{array}\right)^{-1}
  \left(\begin{array}{cc} 1 & 0\\ 0 & -1\end{array}\right)
  \left(\begin{array}{cc} \bar f(\zW) & 1\\ f(\zW) & -1\end{array}\right),
\end{equation}
with the consequence 
\begin{equation}\label{eigen}
 [\Rp(\zW)-\Eins]\left(\begin{array}{c} 1\\ f^+(\zW)\end{array}\right)=0,\quad
 [\Rp(\zW)+\Eins]\left(\begin{array}{c} 1\\ -\bar f^+(\zW)\end{array}\right)=0.
\end{equation}
By definition \eqref{Rpdef}, the elements $R^+_{ik}$ of the matrix $\Rp$ obey the conditions
\begin{equation}
 \bar R^+_{11}=-R^+_{11},\quad \bar R^+_{22}=-R^+_{22},\quad
 \bar R^+_{12}= R^+_{12},\quad \bar R^+_{21}=R^+_{21}.
\end{equation}
Hence the two equations \eqref{eigen} are complex conjugate. $f^+(\zW)$ can now be calculated from \eqref{eigen} provided that $\det(\Rp-1)=0$ holds. By definition, $\det\Rp=-1$, such that $\det(\Rp-1)=-\mathrm{tr}\,\Rp$. Hence,
\begin{equation}\label{equiv}
 \det(\Rp-\Eins)=0\quad\Leftrightarrow\quad\mathrm{tr}\,\Rp=0
\end{equation}
are alternative formulations of the constraints. With the aid of the polynomial matrix $\rp=(r^+_{ik})$,
\begin{equation}\label{rp}
 \rp=\Rp\prod_{l=1}^4(\zW-K_l),
\end{equation}
$f^+(\zW)$ takes the form
\begin{equation}\label{Af}
 f^+(\zW)=\frac{\prod\limits_{l=1}^4(\zW-K_l)-r^+_{11}}{r^+_{12}},\quad
 r^+_{11}+r^+_{22}=0.
\end{equation}
According to \eqref{equiv}, the constraints $r^+_{11}+r^+_{22}=0$ can be reformulated to give 
\begin{equation}\label{zero}
 \left[\prod\limits_{l=1}^4(\zW-K_l)-r^+_{11}\right]
  \left[\prod\limits_{l=1}^4(\zW-K_l)+r^+_{11}\right]=r^+_{12}r^+_{21},
\end{equation}
where by definition the two factors (``brackets'') on the left hand side are normalized \emph{complex conjugate} polynomials of fourth degree ($\bar r^+_{11}=-r^+_{11}$)
whereas $r^+_{12}$ and $r^+_{21}$ are normalized \emph{real} polynomials of fourth degree.

Identifying the zeros on both sides of \eqref{zero} we see see that each factor on the left hand side has to have two zeros of $r^+_{12}$ as well as of $r^+_{21}$. (Note that the brackets are complex conjugate.) Hence the numerator and the denominator of $f^+$ as factors in \eqref{zero} have two common zeros such that $f^+$ has to be a quotient of two polynomials of second degree,
\begin{equation}\label{axisf}
 f^+(\zW)=\frac{n_2(\zW)}{d_2(\zW)}=
 \frac{\zW^2+q\zW+r}{\zW^2+s\zW+t},
\end{equation}
where the complex constants $q$, $r$, $s$, $t$ are restricted by constraints \eqref{Kcond}. Axis values of the form \eqref{axisf} are characteristic for the Ernst potential of the double-Kerr-NUT solution. Before continuing $f^+(\zW)$ to all space, we will discuss the gravitomagnetic potential $a$ on $\Apm$, $\An$, $\mathcal H^{(1/2)}$. For this purpose we express $\chi(-\lambda)$, $\psi(-\lambda)$ in \eqref{aLPalt} by the elements of the last column of $\bPhi$ in \eqref{phiI}. The result
\begin{equation}\label{aLPx}
 a^I=-2\ii K^2\diff{}{K}B^I(K)\big|_{K\to\infty}-C
\end{equation}
tells us that the gravitomagnetic potential has constant values on all intervals $\Apm$, $\An$, $\mathcal H^{(1/2)}$. Since $a(\rW,\zW)$ is only determined up to an arbitrary constant, we adjust $C$ so that
\begin{equation}\label{acon}
 a^+=0.
\end{equation}
To connect the values of $a$ on adjoining intervals we make use of the successive transformations \eqref{cont4} but forgo for the moment the special gauge \eqref{gauge}. Considering the transition $\Ap/\Ha$, we assume that the asymptotics of $\Lp$ can be described by a power series in $1/K$, with
\begin{equation}\label{asymp}
 D^+(K)=1+\mathcal O\left(\frac{1}{K}\right),\quad
 B^+(K)=\mathcal O\left(\frac{1}{K}\right).
\end{equation}
Obviously, our standard gauge \eqref{gauge} satisfies this assumption. Applying 
\eqref{aLPx} to the 1-2 element of the matrix equation
\begin{equation}
 \La=\left(\Eins-\frac{\Fa}{2\ii\Omega^{(1)}(K-K_1)}\right)\Lp,
\end{equation}
see \eqref{cont4}, one obtains 
\begin{equation}\label{atran1}
 a^{(1)}=a^+-\frac{1}{\Omega^{(1)}}.
\end{equation}
Continuing the procedure according to \eqref{cont4} one arrives at
\begin{equation}\label{atran2}
 a^{0}=a^{(1)}+\frac{1}{\Omega^{(2)}},\quad
 a^{(2)}=a^0-\frac{1}{\Omega^{(3)}},\quad
 a^{-}=a^{(2)}+\frac{1}{\Omega^{(4)}}.
\end{equation}
In our context, $\Omega^{(1)}=\Omega^{(2)}=\Omega_1$, $\Omega^{(3)}=\Omega^{(4)}=\Omega_2$. However, the form of Eqs. \eqref{atran1}, \eqref{atran2} remains unchanged repeating their derivation with $\Omega^{(1)}\neq\Omega^{(2)}$, $\Omega^{(3)}\neq\Omega^{(4)}$ in \eqref{cont4}. We shall make use of this later on; see \eqref{omom}.

We can now make sure that our standard representation \eqref{phimin} satisfies the boundary conditions \eqref{B1x} and \eqref{B2x}. Indeed, $L_{12}^+ = B^+(K)=0$ and $L_{22}^+=D(K)=1$ obey assumption \eqref{asymp} such that Eqs. \eqref{atran1}, \eqref{atran2} hold. With $a^+=0$ (see \eqref{acon}) and $\Omega^{(1)}=\Omega^{(2)}=\Omega_1$, $\Omega^{(3)}=\Omega^{(4)}=\Omega_2$ they take the form
\begin{equation}\label{By}
 \begin{aligned}
  \Apm,\An:\quad &a^\pm=0,\quad a^0=0\\
  \mathcal H^{(i)}:\quad &a^{(i)}=-\frac{1}{\Omega_i},\quad i=1,2
 \end{aligned}
\end{equation}
in accordance with \eqref{B1x}, \eqref{B2x}.

\subsection{Equilibrium conditions}
We have shown that the Ernst potential \eqref{Af} can be reduced to a quotient of two polynomials of second degree by means of the constraints $r_{11}+r_{22}=0$.
However, the explicit determination of the coefficients $q$, $r$, $s$, $t$ in \eqref{axisf} turns out to be a subtle point. It depends on a suitable reparametrization of the constants $\Omega_1$, $\Omega_2$, $K_1-K_2$, $K_2-K_3$, $K_3-K_4$, $f_1$, \dots, $f_4$, which permits an easier handling of the constraints. For this reason we introduce the functions
\begin{equation}\label{albet}
 \alpha(\zW)=\frac{\bar d_2(\zW)}{d_2(\zW)},\quad \alpha\bar\alpha=1,\quad
 \beta(\zW)=\frac{\bar n_2(\zW)}{n_2(\zW)},\quad \beta\bar\beta=1
\end{equation}
and discuss their behavior at the points $\zW=K_i$, ($i=1,\dots,4$) which fix the positions of the horizons. To begin with, we examine the configuration of two 
\emph{extended} horizons
\begin{equation}
 \Ha:\quad K_1\ge\zW\ge K_2,\quad
 \Hb:\quad K_3\ge\zW\ge K_4,\quad
 K_1>K_2>K_3>K_4.
\end{equation}
Introducing the parameters
\begin{equation}\label{albet1}
\begin{aligned}
 &\alpha_i=\alpha(K_i),& \alpha_i\bar\alpha_i=1,\quad i=1,\dots,4,\\
 &\beta_i=\beta(K_i),& \beta_i\bar\beta_i=1,\quad i=1,\dots,4,
\end{aligned}
\end{equation}
we obtain from \eqref{albet} the two linear algebraic systems of equations
\begin{equation}\label{lineq}
 \bar d_2(K_i)-\alpha_i d_2(K_i)=0,\quad
 \bar n_2(K_i)-\beta_i n_2(K_i)=0, \quad i=1,\dots,4
\end{equation}
for $s$, $t$ ($\bar s$, $\bar t$); $q$, $r$ ($\bar q$, $\bar r$) with $K_i$, $\alpha_i$, $\beta_i$ as coefficients. According to \eqref{Achsinv}, \eqref{FG} and \eqref{Rpdef} and using $\bar r_{11}^+=-r_{11}^+$ we have
\begin{equation}\label{Up}
 \ee^{2U^+}
 =\frac{(\zW-K_1)(\zW-K_2)(\zW-K_3)(\zW-K_4)}{r_{12}^+}, 
\end{equation}
where $r_{12}^+$ is a real normalized polynomial of fourth degree in $\zW$. From $\ee^{2U^+(K_i)}=0$ ($r_{12}^+(K_i)\neq 0$, $i=1,\dots, 4$) we get
\begin{equation}
 f^+(K_i)=f_i=-\bar f^+(K_i)=-\bar f_i,
\end{equation}
with the consequence
\begin{equation}
 \beta_i=-\alpha_i.
\end{equation}
 Hence, $f^+$ can be expressed in terms of $\alpha_i$ (and $K_i$) alone. Solving the linear equations \eqref{lineq} for $q$, $r$, $s$, $t$ and plugging the result into \eqref{axisf} we arrive at a determinant representation of the axis potential $f^+$ on $\Ap$,
\begin{equation}\label{axisf1}
 f^+(\zW) = \frac{\left|\begin{array}{ccccc}
   1 & K_1^2  & K_2^2  & K_3^2 & K_4^2\\[1ex]
  \ 1\ & \ \alpha_1K_1(\zW-K_1)\ & \ \alpha_2K_2(\zW-K_2)\ &
   \ \alpha_3K_3(\zW-K_3)\ & \ \alpha_4K_4(\zW-K_4)\ \\[1ex]
   0 & K_1    & K_2    & K_3   & K_4\\[1ex]
   0 & \alpha_1(\zW-K_1) & \alpha_2(\zW-K_2) &
   \alpha_3(\zW-K_3) & \alpha_4(\zW-K_4)\\[1ex]
   0 & 1      & 1      & 1     & 1             
   \end{array}\right|}
   {\left|\begin{array}{ccccc}
   1 & K_1^2  & K_2^2  & K_3^2 & K_4^2\\[1ex]
  -1 & \ \alpha_1K_1(\zW-K_1)\ & \ \alpha_2K_2(\zW-K_2)\ &
   \ \alpha_3K_3(\zW-K_3)\ & \ \alpha_4K_4(\zW-K_4)\ \\[1ex]
   0 & K_1    & K_2    & K_3   & K_4\\[1ex]
   0 & \alpha_1(\zW-K_1) & \alpha_2(\zW-K_2) &
   \alpha_3(\zW-K_3) & \alpha_4(\zW-K_4)\\[1ex]
   0 & 1      & 1      & 1     & 1             
   \end{array}\right|}.
\end{equation}
It can easily be seen that
\begin{equation}\label{genf}
 f(\rW,\zW) = \frac{\left|\begin{array}{ccccc}
   1 & K_1^2  & K_2^2  & K_3^2 & K_4^2\\[1ex]
  \ 1\ & \ \alpha_1K_1r_1\ & \ \alpha_2K_2r_2\ &
   \ \alpha_3K_3r_3\ & \ \alpha_4K_4r_4\ \\[1ex]
   0 & K_1    & K_2    & K_3   & K_4\\[1ex]
   0 & \alpha_1r_1 & \alpha_2r_2 &
   \alpha_3r_3 & \alpha_4r_4\\[1ex]
   0 & 1      & 1      & 1     & 1             
   \end{array}\right|}
   {\left|\begin{array}{ccccc}
   1 & K_1^2  & K_2^2  & K_3^2 & K_4^2\\[1ex]
  -1 & \ \alpha_1K_1r_1\ & \ \alpha_2K_2r_2\ &
   \ \alpha_3K_3r_3\ & \ \alpha_4K_4r_4\ \\[1ex]
   0 & K_1    & K_2    & K_3   & K_4\\[1ex]
   0 & \alpha_1r_1 & \alpha_2r_2 &
   \alpha_3r_3 & \alpha_4r_4\\[1ex]
   0 & 1      & 1      & 1     & 1             
   \end{array}\right|},
\end{equation}
where
\begin{equation}\label{ri}
 r_i := \sqrt{(\zW-K_i)^2+\rW^2}\ge0,\qquad i=1,\dots,4,
\end{equation}
is a continuation of $f^+(\zW)$ to all space. (Replace $\zW-K_i=|\zW-K_i|$, $i=1,\dots,4$ in \eqref{axisf1} by $r_i$, $i=1,\dots,4$.)

$f(\rW,\zW)$ is a solution of the Ernst equation \cite{Kramer1980}. As we have already mentioned, the other gravitational potentials $k$, $a$ ($\ee^{2U}=\Re f!$) can be calculated from $f(\rW,\zW)$ via line integrals. This solution of the Einstein vacuum equations is known under the name of double-Kerr-NUT solution. Since it can be shown by the inverse scattering methods that the axis values $f^+(\zW)$ uniquely determine the Ernst potential everywhere in the $\rW$-$\zW$ plane, $f(\rW,\zW)$ as defined in \eqref{genf} is the \emph{only} solution of the Ernst equation to the boundary values \eqref{axisf1}. Hence, \emph{the solution of the two-horizon problem must be a (particular) double-Kerr-NUT solution}.

The double-Kerr-NUT solution itself is a particular case ($N=2$, Minkowski seed) of a class of solutions generated by an $N$-fold B\"acklund transformation from an arbitrary seed solution  of the vacuum field equations \cite{Neugebauer1979,Neugebauer1980a,Neugebauer1980b}\footnote{It should be noted that these solutions can also be written as a quotient of two determinants in complete analogy to \eqref{genf}; see \cite{Kramer1980} or \cite{Kramer}.}. The interrelationship with a class of solitonic solutions discovered by Belinski and Zakharov \cite{BelZak} is discussed in \cite{Kramer}.

Many attempts have been made to establish a connection between the double-Kerr-NUT solution and two-black-hole equilibrium configurations \cite{Kramer1980,Kihara1982,Tomimatsu,Dietz,Hoenselaers1984,Manko2000,Manko2001}. Applying the boundary conditions
\begin{equation}\label{Bz}
 \begin{aligned}
  \Ap,\An:\quad & a=0,\quad k=0\\
  \mathcal C:\quad  & f\to 1,\quad k\to 1
 \end{aligned}
\end{equation}
to $f(\rW,\zW)$, Tomimatsu and Kihara \cite{Kihara1982,Tomimatsu} derived a complete set of algebraic equilibrium conditions on the axis of symmetry connecting the parameters $\alpha_i$, $K_i$ ($i=1,\dots,4$) between each other. Particular solutions of the algebraic system involving numerical results were discussed by Hoenselaers \cite{Hoenselaers1984}, who came to conjecture that the double-Kerr-NUT solution cannot describe the equilibrium between two aligned rotating black holes with positive Komar masses. Hoenselaers and Dietz \cite{Hoenselaers1983,Dietz} and Krenzer \cite{Krenzer} were able to prove this conjecture for symmetric configurations $K_1-K_2=K_3-K_4$, $\Omega_1=\Omega_2$. The explicit solution of the Tomimatsu-Kihara equilibrium conditions was found by Manko et al. \cite{Manko2000,Manko2001}. Finally, Manko and Ruiz \cite{Manko2001} were able to prove Hoenselaers' conjecture. The results derived by Manko and collaborators are important steps toward a non-existence proof. In particular, we will make use of their solution of the equilibrium conditions. Before using these results, we had to formulate and analyze a boundary problem for disconnected horizons, cf. \cite{Varzugin1997}, since a non-existence proof cannot be based on an arbitrarily chosen solution, even though this solution (here: the double-Kerr-NUT solution) seems to be a promising candidate.
Fortunately (or, as expected) the analysis of the boundary problem led to the double-Kerr-NUT solution. 

There is another critical point in the argumentation of Hoenselaers and Dietz and Manko et al. To the best of our knowledge there is no argument in favor of the positiveness of the individual Komar masses of interacting bodies and black holes. On the contrary, Ansorg and Petroff \cite{Ansorg2006} have given convincing counterexamples. We replace the Komar mass inequality (``positivity of the Komar mass of each black hole'') by an inequality connecting angular momentum and horizon area \cite{Hennig2008a}; see Sec.~\ref{insi}. This relation is based on the geometry of trapped surfaces in the interior vicinity of the event horizon.

Our next goal is the reformulation of the constraints \eqref{equiv} in terms of the new parameters $\alpha_i$ ($i=1,\dots,4$). In order to utilize the tools of the inverse (scattering) method we make use of the pseudopotential $\bPhi(\rW,\zW,\lambda)$ as a solution of the LP \eqref{LP}, whose coefficients $N(\rW,\zW)$, $M(\rW,\zW)$ are explicitly known by \eqref{genf} and \eqref{firstin}. The integration of the LP yields (cf. \cite{Neugebauer1996})
\begin{equation}\label{chi}
 \chi(\rW,\zW,\lambda)=\frac{1}{K^2}
 \frac{\left|\begin{array}{ccccc}
        K^2                & K_1^2            & K_2^2    & K_3^2     & K_4^2\\
        \lambda K(K+\ii z) & \alpha_1 K_1 r_1 & \alpha_2 K_2 r_2 & \alpha_3 K_3 r_3 & \alpha_4 K_4 r_4\\
        K & K_1 & K_2 & K_3 & K_4\\
        \lambda (K+\ii z) & \alpha_1 r_1 & \alpha_2 r_2 & \alpha_3 r_3 & \alpha_4 r_4\\
        1 & 1 & 1 & 1 & 1
       \end{array}\right|}{
       \left|\begin{array}{ccccc}
        1                & K_1^2            & K_2^2    & K_3^2     & K_4^2\\
        -1 & \alpha_1 K_1 r_1 & \alpha_2 K_2 r_2 & \alpha_3 K_3 r_3 & \alpha_4 K_4 r_4\\
        0 & K_1 & K_2 & K_3 & K_4\\
        0 & \alpha_1 r_1 & \alpha_2 r_2 & \alpha_3 r_3 & \alpha_4 r_4\\
        0 & 1 & 1 & 1 & 1
       \end{array}\right|},
\end{equation}
where $r_i=\sqrt{(\zW-K_i)^2+\rW^2}\ge0$ and $i=1,\dots,4$. Note that the remaining elements of $\bPhi$ can easily be constructed from $\chi(\rW,\zW,\lambda)=\Phi_{21}$: $\Phi_{11}=\psi(\rW,\zW,\lambda)=\bar\chi(\rW,\zW,\bar\lambda^{-1})$\footnote{Due to \eqref{lambda}, $K$ is a rational function of $\lambda$ and can be replaced in all elements of $\bPhi$ to obtain $\bPhi(\rW,\zW,\lambda)$ as a polynomial of fourth degree in $\lambda$. $\lambda\to1/\bar\lambda$ implies $K\to\bar K$ and vice versa.}, see \eqref{psibar}, $\Phi_{12}=\psi(\rW,\zW,-\lambda)$, $\Phi_{22}=-\chi(\rW,\zW,-\lambda)$. The straightforward verification of \eqref{chi} by inserting $\bPhi(\rW,\zW,\lambda)$ into the LP \eqref{LP} with coefficients \eqref{firstin} and \eqref{genf} is laborious. Instead \cite{Neugebauer1996}, one realizes that $\bPhi_{,z}\bPhi^{-1}$, for fixed values of $\rW$, $\zW$ is regular in the $\lambda$-plane with the exception of a simple pole of first order at $\lambda=\infty$. From Liouville's theorem one may conclude that $\bPhi_{,z}\bPhi^{-1}=\bP+\lambda\bQ$, where the $2\times 2$ matrices $\bP(\rW,\zW)$, $\bQ(\rW,\zW)$ do not depend on $\lambda$. As a consequence of \eqref{uplow}, $\bP$ becomes diagonal and $\bQ$ off-diagonal with $Q_{12}=P_{11}$ and $Q_{21}=P_{22}$ such that one arrives at the first equation \eqref{LP}. From this equation one obtains the second equation \eqref{LP} by using \eqref{psibar}. Note that the Ernst potential \eqref{genf} follows from \eqref{chi}, $f(\rW,\zW)=\chi(\rW,\zW,1)$ ($K\to\infty$) as expected; see \eqref{fLP}. 

Having discussed some implications of the reparametrization for the representation of the Ernst potential and the pseudopotential $\bPhi$, we will now formulate the constraints in terms of $\alpha_i$ and $K_i$, $i=1,\dots,4$. In particular, we will show that representation \eqref{axisf1} of $f^+(\zW)$ together with the boundary conditions \eqref{B1x}, \eqref{B2x} for the gravitomagnetic potential $a$ is equivalent to representation \eqref{Af} with $\mathrm{tr}\,\rp=0$ ($\Leftrightarrow \mathrm{tr}\,\Rp=0$). 

Denote the pseudopotential $\bPhi$ constructed from \eqref{chi} by $\bPhi_\alpha$ and equip all quantities derived from $\bPhi_\alpha$ with an index $\alpha$. Since $\bPhi_\alpha$ is an integral of the LP \eqref{LP}, its values on the axis and horizons have the form \eqref{phiI}. They are continuous at the points of intersection $\Ap/\Ha$, $\Ha/\An$, $\An/\Hb$, $\Hb/\Am$. This may directly be verified in \eqref{chi} and for the other elements of the matrix $\bPhi_\alpha$. The introduction of rotating systems of reference in \ref{inboun} and corotating quantities such as the corotating pseudopotential $\bPhi'$ are closely connected with the definition of the Killing horizon, which has a well-defined angular velocity in contrast to the intervals $[K_1,K_2]$ and $[K_3,K_4]$ as ``potential'' horizons. However, one can introduce different corotating systems of reference at the ends of each ``potential'' horizon with angular velocities $\Omega^{(i)}$, $i=1,\dots,4$, as defined in \eqref{atran1} and \eqref{atran2},
\begin{equation}\label{omom}
 \Omega^{(1)}=\left[a_\alpha^+-a_\alpha^{(1)}\right]^{-1}\!\!,\
 \Omega^{(2)}=\left[a_\alpha^0-a_\alpha^{(1)}\right]^{-1}\!\!,\
 \Omega^{(3)}=\left[a_\alpha^0-a_\alpha^{(2)}\right]^{-1}\!\!,\
 \Omega^{(4)}=\left[a_\alpha^--a_\alpha^{(2)}\right]^{-1}\!\!,
\end{equation}
where, in general, $a_\alpha^+\neq a_\alpha^0\neq a_\alpha^-$. Thus one obtains
\begin{equation}
 \Lp_\alpha\left(\begin{array}{cc}0 & 1\\1 & 0\end{array}\right)(\Lp_\alpha)^{-1}
 =\Rp_\alpha:=\prod\limits_{i=1}^4\left(\Eins-(-1)^i
 \frac{\mathbold{F}_i}{2\ii\Omega^{(i)}(K-K_i)}\right)
 \left(\begin{array}{cc}0 & 1\\1 & 0\end{array}\right),
\end{equation}
where $f_i=f(\rW=0,\zW=K_i)$, $i=1,\dots,4$, entering $\mathbold F_i$ according to \eqref{Fi}, can be taken from \eqref{genf}. A direct consequence of these equations is
\begin{equation}
 \mathrm{tr}\,\Rp_\alpha=0
\end{equation}
which may be compared with the constraints \eqref{Kcond}. Obviously, the constraints $\mathrm{tr}\,\Rp=0$ are satisfied if $\Omega^{(1)}=\Omega^{(2)}$, $\Omega^{(3)}=\Omega^{(4)}$, i.e. if 
\begin{equation}\label{ap}
 \begin{aligned}
 & a^+_\alpha=a^0_\alpha,\quad a^-_\alpha=a^0_\alpha,\\
 & \Omega_1=\Omega^{(1)}=\Omega^{(2)}=\left[a^+_\alpha-a^{(1)}_\alpha\right]^{-1}\!\!,\quad
 \Omega_2=\Omega^{(3)}=\Omega^{(4)}=\left[a^-_\alpha-a^{(2)}_\alpha\right]^{-1}\!\!.
 \end{aligned}
\end{equation}

Conversely, we have shown that the representation \eqref{Af} of $f^+(\zW)$ together with the constraints $\mathrm{tr}\,\rp=0=\mathrm{tr}\,\Rp$ implies conditions \eqref{By}.
Thus we may conclude that \emph{the (four) conditions \eqref{ap} are  a reformulation of the (four) constraints \eqref{traf}}. According to \eqref{B1x}, \eqref{B2x} the conditions \eqref{ap} are necessary conditions for the equilibrium of the two-black-hole configuration. Hence, \emph{constraints \eqref{traf} and restrictions \eqref{ap} are equivalent formulations of the equilibrium conditions to the respective Ernst potentials}. 

The explicit form of the axis equilibrium conditions \eqref{ap} can be evaluated by using \eqref{aLP} or \eqref{aLPx}. In both cases, one may start with expression 
\eqref{chi} that determines all elements of $\bPhi_\alpha$ ($\ee^{2U}=\Re f$ in \eqref{aLP} can be taken from \eqref{genf}). Equation \eqref{aLP} yields the gravitomagnetic potential $a(\rW,\zW)$ everywhere including the axis intervals $I$ whereas equation \eqref{aLPx} directly leads to the axis values of $a^I$,\footnote{Henceforth we omit the index $\alpha$.} $I=\Apm,\An,\mathcal H^{(1/2)}$ ($B^I(K)$ may be taken from representation \eqref{phiI}). Straightforward calculations result in
\begin{equation}\label{aI}
 a^I+C=-\frac{\ii}{H^I}\left|
       \begin{array}{cccc}
        K_1^2 & K_2^2 & K_3^2 & K_4^2\\
        K_1   & K_2   & K_3   & K_4\\
        1     & 1     & 1     & 1\\
        \pi^I_1 & \pi^I_2 & \pi^I_3 & \pi^I_4
       \end{array}\right|,
\end{equation}
where
\begin{equation}\label{aIx}
 H^I=\left|\begin{array}{cccc}
            \lambda_1^I\alpha_1 K_1 & \lambda_2^I\alpha_2 K_2 & \lambda_3^I\alpha_3 K_3 & \lambda_4^I\alpha_4 K_4\\
            K_1 & K_2 & K_3 & K_4\\
            \lambda_1^I\alpha_1 & \lambda_2^I\alpha_2 & \lambda_3^I\alpha_3 & \lambda_4^I\alpha_4\\
            1 & 1 & 1 & 1
           \end{array}\right|
\end{equation}
with
\begin{equation}\label{pi}
 \pi^I_k=\frac{\prod\limits_{n=1}^4(1+\lambda^I_n\alpha_n)}{1+\lambda^I_k\alpha_k},\quad \lambda^I_k=\frac{|K_k-\zW|}{K_k-\zW},\quad I=\Apm,\An,\mathcal H^{(1,2)},\quad i=1,\dots,4.
\end{equation}
Note that $\lambda_k^I=\pm 1$ marks the interval $I$, e.g., $I=\An$, $K_4\le K_3\le\zW\le K_2\le K_1$: $(\lambda_k^0)=(1,1,-1,-1)$.

Inserting determinant expressions \eqref{aI} in the axis equilibrium conditions \eqref{ap}\footnote{We omit $\alpha$ as arranged.} one obtains the angular velocities $\Omega_1$, $\Omega_2$ of the horizons $\mathcal H^{(1/2)}$ in terms of the parameters $\alpha_i$, $K_i$, $i=1,\dots,4$ and two conditions ($a^+=a_0$, $a^-=a^0$) that restrict the choice of these parameters. These restrictions must be taken into account when examining the Ernst potential \eqref{genf}.

\subsection{Discussion of the solution}
We have shown that the solution of the boundary problem \eqref{B1x}-\eqref{B3x} for the Ernst equation \eqref{Ernst} is given by the Ernst potential \eqref{genf}, whose constant parameters $\alpha_i$, $K_i$, $i=1,\dots,4$ have to satisfy the first set of the axis equilibrium conditions \eqref{ap}, $a^+=a^0$, $a^-=a^0$ with $a^\pm$, $a^0$ from \eqref{aI}. ($\Omega_1$, $\Omega_2$ can be calculated straightforwardly.) The appearance of the equilibrium condition is not promising and makes a comprehensive discussion of the solution difficult. We will only list a few aspects of the interpretation.

\subsubsection{Number of parameters}
Written in dimensionless coordinates such as
\begin{equation}
 \tilde\rW=\frac{\rW}{K_{23}},\quad \tilde\zW=\frac{\zW-K_1}{K_{23}},\quad
 K_{23}=K_2-K_3,
\end{equation}
the Ernst potential contains four free parameters. Since the quotient of determinants \eqref{genf} remains unchanged under a translation of the two-black-hole configuration along the $\zW$-axis and a multiplication of $\rW$, $\zW$ and $K_i$ by a common (real) factor, $f$ depends on the coordinates $\tilde\rW$, $\tilde\zW$ and the six parameters $\alpha_1,\dots,\alpha_4$; $K_{12}/K_{23}$, $K_{34}/K_{23}$, where we have used the abbreviation
\begin{equation}
 K_{ij}=K_i-K_j,\quad i,j=1,\dots,4.
\end{equation}
It can easily be seen that the two conditions $a^+=a^0$, $a^-=a^0$ can be rescaled to be written in terms of the six parameters alone. Hence $f=f(\tilde\rW,\tilde\zW)$ is a four parameter solution.

\subsubsection{Singularities outside the horizons}
The solution is a necessary consequence of the integration of the LP along the boundary (closed dashed line in Fig.~\ref{Fig1}). We have (implicitly) assumed the validity of the Ernst equation everywhere in the enclosed $\rW$-$\zW$ domain. It is, however, not clear, whether the Ernst potential is really free of singularities there. As matters stand at present the question must remain undecided. Interestingly, the solution of the static two-horizon problem on the level of the Laplace equation $\Delta U=0$ (which is the static form of the Ernst equation) has no singularities outside the horizons. (Applying the inverse formalism one simply obtains a superposition of two Schwarzschild solutions in Weyl-Lewis-Papapetrou coordinates). The so-called ``conical'' singularity of the metric on $\An$ that forbids the existence of static two-black-hole configurations only appears when the metric coefficient $\ee^{2k}$ is involved. Calculated from a \emph{regular solution} of the Laplace equation (the ``double-Schwarzschild'' solution), $\ee^{2k}$ violates the regularity condition $\ee^{2k}=1$ on $\An$; see \eqref{B1}. We will discuss this condition  (which ensures elementary flatness on the axis of symmetry) in the next section.

\subsubsection{Generalizations}
Obviously, our analysis of the two-horizon problem can easily be extended to an arbitrary number $n$, $n>2$, of aligned disconnected horizons. Integrating the LP \eqref{LP} along the $2n+1$ intervals $I$ ($n$ horizons $\mathcal H$ and $n+1$ ``regular'' intervals $\mathcal A$) one arrives at a representation of the form \eqref{phiI} for each of the $2n+1$ intervals. Consequently, one has $2n+1$ matrices $\mathbold{L}^I$ and $2n$  matrices $\mathbold{F}_i$; cf.~\eqref{Fi}. Replacing the symbol $\prod_{i=1}^4$ in \eqref{Rpdef} by $\prod_{i=1}^{2n}$ one obtains, via \eqref{FG} and by \eqref{Achsf}, a representation for the Ernst potential $f^+(\zW)$ on $\Ap$. The rational structure of this potential ($f^+$ is a quotient of two normalized polynomials of equal degree in $\zW$) is a characteristic feature of solutions to the Einstein equations derived by iterative B\"acklund transformations of the metric of the Minkowski space \cite{Kramer}, \cite{Neugebauer1980a}, or, equivalently, by the Belinski-Zakharov approach \cite{BelZak}. This confirms Varzugin's result \cite{Varzugin1997} which says that any equilibrium configuration of aligned black holes can be described by a Belinski-Zakharov solution \cite{BelZak}. After a reparametrization in full analogy to \eqref{albet}, $f(\rW,\zW)$ again turns out to be a quotient of $(2n+1)\times(2n+1)$ determinants whose structure is an obvious generalization of the determinants in \eqref{genf}. Finally, the equilibrium conditions can be derived straightforwardly.

\section{Elementary flatness and equilibrium conditions}
We know from the Bach-Weyl paper \cite{Bach} that the metric coefficient $\ee^{2k}$ is a measure for the interaction of the two black holes. To guarantee equilibrium, $k$ has to vanish on the portion of the axis of symmetry between the two black holes. From a geometrical point of view, the condition $\ee^{2k}=1$ on $\mathcal A$ is a necessary condition for elementary flatness (Lorentzian geometry) of spacetime in the vicinity of the rotation axis $\mathcal A$. Our discussion of the metric potential $k=k(\rW,\zW)$ is based on Kramer's representation \cite{Kramer1986}, which is a result of the integration of the defining relation \eqref{k1} with $f(\rW,\zW)$ from \eqref{genf}. It turns out that $k(\rW=0,\zW)$ is a step function with constant values on the intervals $\Apm$, $\An$. In particular, one has $\ee^{2k^+}=\ee^{2k^-}$. Equation \eqref{k1} determines $k$ up to an additive constant, which can be chosen such that $k^+=0$ with the consequence $\ee^{2k^-}=1$. For this choice, $\ee^{2k^0}$ takes the form
\begin{equation}
 \ee^{2k^0}=1+\left(\frac{H^0}{H^+}-1\right)\left(\frac{H^0}{H^+}+1\right)
\end{equation}
with $H^I$, $I=\An,\Ap$ as in \eqref{aIx}. Obviously, the equilibrium condition $\ee^{2k^0}=1$ has two solution branches, $H^0=\pm H^+$. Tomimatsu and Kihara \cite{Tomimatsu} ruled out the condition $H^0=H^+$ ($H^0=H^+$ together with $a^+=a^-$, $a^-=a^0$ leads to overlapping horizons). $H^0=-H^+$ yields
\begin{equation}\label{kBed}
 \alpha_1\alpha_2+\alpha_3\alpha_4=0
\end{equation}
and effects a considerable simplification of the equilibrium conditions \eqref{ap}. Putting \eqref{aI}-\eqref{pi} in $a^+=a^0$, $a^-=a^0$ and observing $H^+=-H^0$ one finds 
\begin{equation}\label{E1}
 \begin{aligned}
  &\alpha_3(1-\alpha_4)^2(K_{41}+K_{32})-(1-\alpha_3\alpha_4)\alpha_{43}K_{31}K_{32}=0\\
  &\alpha_1(1+\alpha_2)^2(K_{41}+K_{32})+(1-\alpha_1\alpha_2)\alpha_{21}K_{31}K_{41}=0,
 \end{aligned}
\end{equation}
where
\begin{equation}
 K_{ij}=K_i-K_j,\quad
 \alpha_{ij}=\frac{\alpha_i-\alpha_j}{K_i-K_j}.
\end{equation}

For \emph{point-like} horizons $K_1=K_2$ or/and $K_3=K_4$, the parametrization \eqref{albet1} does not apply, since the mapping of the four coefficients $q$, $r$, $s$, $t$ in \eqref{axisf} onto less than four parameters $\alpha_i$ ($\alpha_1=\alpha_2$ or/and $\alpha_3=\alpha_4$) is not invertible. However, the invertibility can be restored by introducing the \emph{derivatives} $\alpha'(\zW)$, $\beta'(\zW)$ in the confluent points $K_1=K_2$ or/and $K_3=K_4$. As was shown in \cite{Hennig2011} this concept makes it possible to consider the equilibrium conditions for configurations with degenerate horizons as particular cases of \eqref{kBed}, \eqref{E1}: To describe point-like configurations one has to set
\begin{equation}
 \begin{aligned}
  &K_2=K_1,\quad \alpha_2=\alpha_1,\quad\alpha_{21}=\alpha'(K_1)=-\ii\alpha_1\gamma_1\quad\textrm{or/and}\\
  &K_4=K_3,\quad \alpha_4=\alpha_3,\quad\alpha_{43}=\alpha'(K_3)=-\ii\alpha_3\gamma_3,
 \end{aligned}
\end{equation}
where $\gamma_1$ and $\gamma_3$ are real constants. 

To introduce $\alpha_{21}$ and $\alpha_{43}$ in the Ernst potential \eqref{genf}, one has to subtract the second column from the third one and the fourth column from the fifth one and to decompose the $\alpha$-parameters into symmetric and antisymmetric parts (e.g. $\alpha_1=\frac{1}{2}(\alpha_1+\alpha_2)+\frac{1}{2}(\alpha_1-\alpha_2)$, $\alpha_2=\frac{1}{2}(\alpha_1+\alpha_2)-\frac{1}{2}(\alpha_1-\alpha_2)$).

We are now prepared to discuss all constellations (extended/extended, extended/point-like, point-like/point-like) by using \eqref{genf}, \eqref{kBed} and \eqref{E1}.

\section{Black hole inequalities and singularities\label{insi}}

\subsection{Two sub-extremal black holes}
We start the discussion of the different types of possible two-black-hole equilibrium configurations by considering spacetimes containing two black holes with extended horizons. Following Both and Fairhurst \cite{Booth}, we will assume that a physically reasonable non-degenerate black hole should be \emph{sub-extremal}, i.e. characterized through the existence of trapped surfaces (surfaces with a negative expansion of outgoing null geodesics) in every sufficiently small interior neighborhood of the event horizon. As shown in \cite{Hennig2008a}, the presence of trapped surfaces implies the inequality $ 8\pi|J|<A$ between angular momentum $J$ and horizon area $A$ of the black hole. In a regular spacetime with two sub-extremal black holes, both black holes have to satisfy this inequality individually,
\begin{equation}\label{ineq}
 8\pi|J_i|<A_i,\quad i=1,2.
\end{equation}
This is the key ingredient for the non-existence proof of two-black-hole equilibrium configurations, as we will see below.

In order to test these inequalities for that subclass of the double-Kerr-NUT family of solutions which describes two gravitating objects with extended horizons, we have to solve the general axis regularity conditions \eqref{kBed}, \eqref{E1}. For $K_1>K_2>K_3>K_4$ these conditions can be written as
\begin{equation}\label{Bed1}
 \alpha_1\alpha_2+\alpha_3\alpha_4=0
\end{equation}
and
\begin{equation}\label{Bed2}
\begin{aligned}
 \frac{(1-\alpha_4)^2}{\alpha_4}w^2&=\frac{(1-\alpha_3)^2}{\alpha_3},\quad
   &w&:=\sqrt{\frac{K_{14}K_{24}}{K_{13}K_{23}}}\in[1,\infty),\\ 
 \frac{(1+\alpha_2)^2}{\alpha_2}w'^2&=\frac{(1+\alpha_1)^2}{\alpha_1},\quad
   &w'&:=\sqrt{\frac{K_{23}K_{24}}{K_{13}K_{14}}}\in(0,1].
\end{aligned}
\end{equation}
As shown by Manko et al. \cite{Manko2001}, these equations can be explicitly solved for $\alpha_1,\dots,\alpha_4$,
\begin{equation}\label{alpha1}
 \begin{aligned}
  &\alpha_1 = \frac{w'\alpha^2+\ii\eps\alpha}{w'-\ii\eps\alpha},\quad
  &&\alpha_2 = \frac{\alpha^2+\ii w'\eps\alpha}{1-\ii w'\eps\alpha}\\
  &\alpha_3 = \frac{w\alpha^2-\alpha}{w-\alpha},\quad
  &&\alpha_4 = \frac{\alpha^2-w\alpha}{1-w\alpha},
 \end{aligned}
\end{equation}
where $\alpha:=\sqrt{-\alpha_1\alpha_2}=\sqrt{\alpha_3\alpha_4}=:\ee^{\ii\phi}$, $\alpha\bar\alpha=1$ and $\eps=\pm1$. 

 Now we are in a position to calculate the quantities
\begin{equation}
p_i:=\frac{8\pi J_i}{A_i},\qquad i=1,2,
\end{equation}
in terms of the parameters $w$, $w'$ and $\phi\in[0,2\pi)$. The result is
\begin{equation}\label{pJ}
 p_1 = \eps\frac{1+\Phi w'}{w'(\Phi+w')},\quad
 p_2 = \eps\frac{w(w-\Phi)}{1-w\Phi}
\end{equation}
with
\begin{equation}
 \Phi:=\cos\phi+\eps\sin\phi,\quad \eps=\pm1.
\end{equation}
Rewriting the inequalities in \eqref{ineq} as $p_i^2<1$, $i=1,2$, we find with the previous formulae the two conditions
\begin{equation}
w'^2+2\Phi w'+1<0\quad\textrm{and}\quad
w^2-2\Phi w+1<0.
\end{equation}
Since the latter inequalities imply $\Phi w'<0$ and $\Phi w>0$, we arrive at a
contradiction, because $w$ and $w'$ are positive. In this way, we have shown that two sub-extremal black holes cannot be in equilibrium.

\subsection{One sub-extremal and one degenerate black hole}

Without loss of generality, we may assume that the upper horizon is
point-like ($K_1=K_2$) and the lower one is extended ($K_2>K_3>K_4$) as sketched in  Fig.~\ref{Fig1}. Then, the equilibrium conditions \eqref{kBed}, \eqref{E1} can be written as
\begin{equation}\label{al1}
 \alpha_1^2+\alpha_3\alpha_4=0,
\end{equation}
with the solution
\begin{equation}
 \alpha_1=\ii\eps\alpha,\quad
 \alpha_3\alpha_4=\alpha^2,
\end{equation}
where $\eps=\pm1$ and $\alpha\in\C$, $\alpha\bar\alpha=1$, and
\begin{equation}\label{al4}
 (1-\alpha_4)w=\frac{\alpha^2-\alpha_4}{\alpha},\quad
 (\alpha-\ii\eps)\left[\gamma_1 K_{23}w(\eps\alpha+\ii)+\ii(w+1)(\eps\alpha-\ii)\right]=0.
\end{equation}
As solution to these equations we obtain \emph{two} sets of parameters. In the first solution branch, the parameters $\alpha_1$,
$\gamma_1$, $\alpha_3$ and $\alpha_4$ have to be chosen according to
\begin{equation}\label{L1}
 \alpha_1=\ii\eps\alpha,\quad
 \gamma_1=\frac{\ii(w+1)}{w K_{23}}\frac{\ii-\eps\alpha}{\ii+\eps\alpha},\quad
 \alpha_3=\frac{w\alpha^2-\alpha}{w-\alpha},\quad
 \alpha_4=\frac{\alpha^2-w\alpha}{1-w\alpha},
\end{equation}
which is the limit $K_1\to K_2$ ($\Leftrightarrow
w'\to1$) of solution \eqref{alpha1} for extended horizons.
This family of
solutions depends on the two parameters $\alpha$ and $w$ (and
on two additional scaling parameters, e.g. $K_1$ and $K_{23}$).

The second solution branch of the equilibrium conditions is given by
\begin{equation}\label{L2}
 \alpha_1=-1,\quad
 \gamma_1\in\R,\quad
 \alpha_3=\frac{1-\ii\eps w}{1+\ii\eps w},\quad
 \alpha_4=-\frac{1+\ii\eps w}{1-\ii\eps w},
\end{equation}
i.e. the corresponding Ernst potential depends
on the two parameters $\gamma_1$ and $w$
(plus two scaling parameters).
Interestingly, this solution has no counterpart in the case of extended
horizons.

 The desired non-existence proof follows the same idea for both families: The ADM mass $M$ of the spacetime can be expressed in terms of the two parameters (and the additional scaling parameter $K_{23}$). The resulting expressions can be estimated, using the inequality $8\pi|J|<A$ for the sub-extremal object. We obtain
\begin{equation}
 M<0
\end{equation}
in contradiction to the positive mass theorem. This indicates the presence of unphysical singularities and we conclude that configurations with one degenerate and one sub-extremal black hole cannot be in equilibrium.

\subsection{Two degenerate black holes}
In the case of possible equilibrium configurations with two degenerate black holes\footnote{See also \cite{Cabrera} for a discussion of properties of spacetimes with two degenerate objects.}, we find three one-parametric families of candidate solutions:
The equilibrium conditions are now
\begin{equation}\label{al2}
 \alpha_1^2+\alpha_3^2=0,
\end{equation}
which is solved by
\begin{equation}
 \alpha_1=\ii\eps\alpha,\quad
 \alpha_3=-\alpha,
\end{equation}
with $\eps=\pm1$ and $\alpha\in\C$, $\alpha\bar\alpha=1$, and 
\begin{equation}\label{al3}
 (\alpha+1)\left[(\alpha-1)\gamma_3K_{23}-2\ii(\alpha+1)\right]=0,\quad
 (\alpha-\ii\eps)\left[\gamma_1 K_{23}(\eps\alpha+\ii)+2\ii(\eps\alpha-\ii)\right]=0.
\end{equation}

Equation \eqref{al3} has three different solutions.
The first one is
\begin{equation}\label{L3}
 \alpha_1=\ii\eps\alpha,\quad \alpha_3=-\alpha,\quad
 \gamma_1=\frac{2\ii}{K_{23}}\cdot\frac{1+\ii\eps\alpha}{1-\ii\eps\alpha},
 \quad
 \gamma_3=\frac{2\ii}{K_{23}}\cdot\frac{\alpha+1}{\alpha-1}
\end{equation}
and depends on one free parameter $\alpha$. This solution can be obtained in the limit
$K_1\to K_2$, $K_3\to K_4$ ($\Leftrightarrow w\to1,w'\to 1$) from
\eqref{alpha1}.

The second and third solution branches are
\begin{equation}\label{L4}
 \alpha_1=-\ii\eps,\quad \alpha_3=1, \quad
 \gamma_1=\frac{2\eps}{K_{23}},
 \quad
 \gamma_3\in\R,
\end{equation}
and
\begin{equation}\label{L5}
 \alpha_1=-1,\quad \alpha_3=-\ii\eps,\quad \gamma_1\in\R,\quad
 \gamma_3=\frac{2\eps}{ K_{23}},
\end{equation}
where now $\gamma_3$ or $\gamma_1$ are free parameters.

In order to show that the above solutions do not lead to regular two-black-hole configurations, we calculate the ADM mass $M$. For the first solution branch, the result is
\begin{equation}
 M=-\frac{2K_{23}}{3+\sqrt{2}\cos
    \left(\eps\phi+\frac{\pi}{4}\right)}.
\end{equation}
Obviously, $M$ is always negative and we arrive again at a contradiction to the positive mass theorem.

For the second and third solution branches, $M$ has the form
\begin{equation}
 M=-\frac{K_{23}}{2}\cdot\frac{\tilde\gamma_3-2}{\tilde\gamma_3-1},\quad
 \tilde\gamma_3=\eps\gamma_3 K_{23}.
\end{equation}
Hence, the mass is negative for $\tilde\gamma_3<1$ and for $\tilde\gamma_3>2$ --- again a contradiction. However, $M>0$ holds in the parameter range $\tilde\gamma\in[1,2]$. In that case it is not difficult to show that the solutions violate the Penrose inequality $M>\sqrt{(A_1+A_2)/16\pi}$. Since this inequality is related to cosmic censorship, this might indicate that these configurations are not regular outside the two gravitational sources. However, since so far no rigorous proof of the Penrose inequality for axisymmetric configurations was given, this does not yet exclude the possibility that two sub-extremal black holes can be in equilibrium. Instead, one can directly show that these solutions always suffer from the presence of singular rings; see Appendix A in \cite{Hennig2011}. Therefore, this solution branch can be excluded, too.
\section{Summary}

As a characteristic example for the present discussion about existence or non-existence of stationary equilibrium configurations within the theory of general relativity, we have studied the question whether \emph{two aligned black holes} can be in equilibrium. We have shown how this question can be reformulated in terms of a boundary value problem for two disconnected Killing horizons with specific boundary conditions at the horizons, on the axis of symmetry and at infinity. Using the Ernst formulation of the Einstein equations, it was possible to apply methods from soliton theory which allow us to study an associated linear problem (LP) instead of the non-linear field equations themselves. 

By integrating this LP along the boundaries of the vacuum region (along the axis, the horizons and at infinity) we found an explicit expression for the Ernst potential $f^+$ on the upper part of the symmetry axis $\Ap$ (see Fig.~\ref{Fig1}). In particular, it turned out that $f^+$ is a quotient of two normalized polynomials in $\zW$ (the ``axis coordinate''), which depends on a number of parameters. In addition, the solution of the LP led to constraints between these parameters which we have shown to be equivalent to the ``equilibrium condition'' (regularity condition) $a=0$ ($a$: gravitomagnetic potential) on the axis of symmetry. By analyzing the constraints, we found that the Ernst potential $f^+$ is a quotient of two polynomials of second degree in $\zW$ and hence all possible equilibrium configurations with two black holes would necessarily belong to the double-Kerr-NUT family of solutions. 

Discussing the three possible configurations
\begin{enumerate}
 \item[(i)] two sub-extremal black holes
 \item[(ii)] one sub-extremal and one degenerate black hole
 \item[(iii)] two degenerate black holes
\end{enumerate}
which are characterized by specific restrictions of the parameters (``constraints'') of the double-Kerr-NUT solutions, we could show that none of them does correspond to a physically reasonable black hole solution: All solution families (i), (ii) and (iii) suffer from the presence of naked singularities, which manifests in the violation of physical black hole inequalities (the positivity of the ADM mass or inequality \eqref{ineq} between angular momentum and horizon areas for sub-extremal black holes). Hence we can conclude that \emph{two-black hole equilibrium configurations do not exist}.
\begin{acknowledgments}
 We would like to thank Reinhard Meinel and Marcus Ansorg for interesting discussions and Robert Thompson for commenting on the manuscript.
\end{acknowledgments}



\begin{thebibliography}{}

\bibitem{Ansorg2006}
Ansorg, M. and Petroff, D.:
Negative Komar mass of single objects in regular, asymptotically flat
spacetimes.
Class. Quantum Grav. {\bf 23}, L81 (2006)
\bibitem{Bach}
Bach, R. and Weyl, H.: Neue L\"osungen der
Einsteinschen Gravitationsgleichungen.
Mathemat. Z. {\bf 13}, 134 (1922)
\bibitem{Beig1}
Beig, R. and Schoen, R. M.:
On static $n$-body configurations in relativity.
Class. Quantum Grav. {\bf 26}, 075014 (2009)
\bibitem{Beig2}
Beig, R., Gibbons G. W. and Schoen R. M.:
Gravitating opposites attract.
Class. Quantum Grav. {\bf 26}, 225013 (2009)
\bibitem{BelZak}
Belinski, V. A. and Zakharov, V. E.:
Integration of the Einstein equations by the inverse scattering method and calculation of exact soliton solutions (in Russian).
Sov. Phys. JETP {\bf 75}, 1953 (1978); Belinski, V. A. and Zakharov, V. E.: Stationary gravitational solutions with axial symmetry (in Russian), Sov. Phys. JETP {\bf 77}, 3 (1979).
\bibitem{Booth}
Booth, I. and Fairhurst, S.:
Extremality conditions for isolated and dynamical horizons.
Phys. Rev. D {\bf 77}, 084005 (2008)
\bibitem{Cabrera}
Cabrera-Munguia, I., Manko, V. S. and Ruiz, E.:
Remarks on the mass-angular momentum relations for two extreme Kerr sources in equilibrium.
Phys. Rev. D {\bf 82}, 124042 (2010)
\bibitem{Carter}
Carter, B.: Black hole equilibrium states in \emph{Black Holes} (Les
Houches), edited by C. deWitt and B. deWitt, Gordon and Breach, London
(1973)
\bibitem{Dietz}
Dietz, W. and Hoenselaers, C.:
Two mass solution of Einstein's vacuum equations: The double Kerr solution.
Ann. Phys. {\bf 165}, 319 (1985)
\bibitem{Ernst}
Ernst, F. J.:
New formulation of the axially symmetric gravitational field
problem II, 
Phys. Rev. {\bf 168} (1968), 1415
\bibitem{Hennig2008a}
Hennig, J., Ansorg, M., and Cederbaum, C.:
A universal inequality between the angular momentum and horizon area for
axisymmetric and stationary black holes with surrounding matter. 
Class. Quantum Grav. {\bf 25}, 162002 (2008)
\bibitem{Hennig2011}
Hennig, J. and Neugebauer, G.: Non-existence of stationary two-black-hole configurations: The degenerate case. arXiv:1103.5248 (2011)
\bibitem{Hoenselaers1983}
Hoenselaers, C. and Dietz, W.:
Talk given at the GR10 meeting, Padova (1983)
\bibitem{Hoenselaers1984}
Hoenselaers, C.:
Remarks on the Double-Kerr-Solution.
Prog. Theor. Phys. {\bf 72}, 761 (1984)
\bibitem{Kihara1982}
Kihara, M. and Tomimatsu, A.:
Some properties of the symmetry axis in a superposition of two Kerr solutions.
Prog. Theor. Phys. {\bf 67}, 349 (1982)
\bibitem{Kramer}
Kramer, D., Stephani, H., Herlt, E. and MacCallum, M.: 
Exact Solutions of Einstein's Field Equations 
Cambridge University Press, Cambridge (1980), p. 335
\bibitem{Kramer1980}
Kramer, D. and Neugebauer, G.: The superposition of two Kerr solutions.
Phys. Lett. A {\bf 75}, 259 (1980)
\bibitem{Kramer1986}
Kramer, D.:
Two Kerr-NUT constituents in equilibrium.
Gen. Relativ. Gravit. {\bf 18}, 497 (1986)
\bibitem{Krenzer}
Krenzer, G.: Schwarze L\"ocher als Randwertprobleme der
axislsymmetrisch-station\"aren Einstein-Gleichungen.
PhD Thesis, University of Jena (2000) 
\bibitem{Manko2000}
Manko, V. S., Ruiz, E., and Sanabria-G\'omez, J. D.:
Extended multi-soliton solutions of the Einstein field equations:
II. Two comments on the existence of equilibrium states.
Class. Quantum Grav. {\bf 17}, 3881 (2000)
\bibitem{Manko2001}
Manko, V. S. and Ruiz, E.:
Exact solution of the double-Kerr equilibrium problem.
Class. Quantum Grav.~{\bf 18}, L11 (2001)
\bibitem{Meinel2008}
Meinel, R., Ansorg, M., Kleinw\"achter, A., Neugebauer, G.
and Petroff, D.:
Relativistic figures of equilibrium,
Cambridge University Press, Cambridge (2008)
\bibitem{Neugebauer1979}
Neugebauer, G.:
B\"acklund transformations of axially symmetric stationary gravitational
fields.
J.~Phys.~A {\bf 12}, L67 (1979)
\bibitem{Neugebauer1980a}
Neugebauer, G.:
A general integral of the axially symmetric stationary Einstein equations.
J.~Phys.~A {\bf 13}, L19 (1980)
\bibitem{Neugebauer1980b}
Neugebauer, G.:
Recursive calculation of axially symmetric stationary Einstein fields.
J.~Phys.~A {\bf 13}, 1737 (1980)
\bibitem{Neugebauer1996}
Neugebauer, G.: Gravitostatics and rotating bodies in \emph{Proc. 46th Scottish Universities Summer School in Physics (Aberdeen)}, edited by G. S. Hall and J. R. Pulham, Copublished by SUSSP Publications, Edinburgh, and Institute of Physics Publishing, London (1996)
\bibitem{Neugebauer2000}
Neugebauer, G.: Rotating bodies as boundary value problems.
Ann. Phys. (Leipzig) {\bf 9}, 342 (2000)
\bibitem{Neugebauer2003}
Neugebauer, G. and Meinel, R.:
Progress in relativistic gravitational theory using the inverse
scattering method.
J. Math. Phys. {\bf 44}, 3407 (2003)
\bibitem{Neugebauer2009}
Neugebauer, G. and Hennig, J.:
Non-existence of stationary two-black-hole configurations.
Gen. Relativ. Gravit. {\bf 41}, 2113 (2009)
\bibitem{Tomimatsu}
Tomimatsu, A. and Kihara, M.:
Conditions for regularity on the symmetry axis in a superposition of two
Kerr-NUT solutions.
Prog. Theor. Phys. {\bf 67}, 1406 (1982)
\bibitem{Varzugin1997}
Varzugin, G. G.: Equilibrium configuration of black holes and the inverse scattering method.
Theor. Math. Phys. {\bf 111}, 667 (1997)
\bibitem{Varzugin1998}
Varzugin, G. G.: The interaction force between rotating black holes in equilibrium. Theor. Math. Phys. {\bf 116}, 1024 (1998)
\bibitem{Yamazaki}
Yamazaki, M.:
Stationary line of $N$ Kerr masses kept apart by gravitational spin-spin
interaction.
Phys. Rev. Lett. {\bf 50}, 1027 (1983)

\end{thebibliography}
\end{document}